\documentclass[12pt]{article}
\usepackage{graphicx}

\title{Quantum-enhanced belief propagation for LDPC decoding}
\usepackage{authblk}

\author[1,2]{Sheila M. Perez-Garcia\footnote{sheila.garcia.17@ucl.ac.uk}}
\affil[1]{Phasecraft Ltd.}
\affil[2]{University College London}
\author[1,3]{Ashley Montanaro\footnote{ashley@phasecraft.io}}
\affil[3]{University of Bristol}
\date{}

\usepackage[a4paper, total={6.25in, 8.5in}]{geometry}
\usepackage[utf8]{inputenc}
\usepackage[T1]{fontenc}
\usepackage{amsmath}
\usepackage{svg}
\usepackage[amsmath]{ntheorem}
\usepackage{amsfonts}
\usepackage{amssymb}
\usepackage{siunitx}
\usepackage{comment}
\AtBeginDocument{\RenewCommandCopy\qty\SI}
\usepackage[toc,page]{appendix}
\usepackage[version=4]{mhchem}
\usepackage{bbold}
\usepackage{algorithm2e}
\RestyleAlgo{ruled}

\definecolor{matred}{rgb}{0.8392156862745098, 0.15294117647058825, 0.1568627450980392}
\definecolor{matorange}{rgb}{1.0, 0.4980392156862745, 0.054901960784313725}
\definecolor{matgreen}{rgb}{0.17254901960784313, 0.6274509803921569, 0.17254901960784313}
\definecolor{matblue}{rgb}{0.12156862745098039, 0.4666666666666667, 0.7058823529411765}
\usepackage{subcaption}
\usepackage[toc,page]{appendix}
\usepackage[style=numeric-comp,natbib=true,sorting=none]{biblatex}
\usepackage{tikz}

\newcommand{\ket}[1]{\lvert#1\rangle}
\newcommand{\bra}[1]{\langle#1\rvert}
\newcommand{\abs}[1]{\lvert#1\rvert}

\usepackage{hyperref}
\hypersetup{
    colorlinks=true,
    linkcolor=black,
    filecolor=magenta,      
    urlcolor=cyan,
    citecolor=blue
}
\usepackage{cleveref}
\newtheorem{lemma}{Lemma}
\crefname{lemma}{lemma}{lemmas}
\Crefname{lemma}{Lemma}{Lemmas}
\begin{document}

\maketitle

\begin{abstract}
    Decoding low-density parity-check codes is critical in many current technologies, such as fifth-generation (5G) wireless networks and satellite communications. The belief propagation algorithm allows for fast decoding due to the low density of these codes. However, there is scope for improvement to this algorithm both in terms of its computational cost when decoding large codes and its error-correcting abilities. Here, we introduce the quantum-enhanced belief propagation (QEBP) algorithm, in which the Quantum Approximate Optimization Algorithm (QAOA) acts as a pre-processing step to belief propagation. We perform exact simulations of syndrome decoding with QAOA, whose result guides the belief propagation algorithm, leading to faster convergence and a lower block error rate (BLER). In addition, through the repetition code, we study the possibility of having shared variational parameters between syndromes and, in this case, code lengths. We obtain a unique pair of variational parameters for level-1 QAOA by optimizing the probability of successful decoding through a transfer matrix method. Then, using these parameters, we compare the scaling of different QAOA post-processing techniques with code length.
    %173 words
\end{abstract}

% \tableofcontents

\section{Introduction}

Low-density parity check (LDPC) codes are a vital component in fifth-generation (5G) new radio (NR) digital communications, replacing turbo codes for forward error correction. This class of linear codes, first proposed by Gallager in his 1963 PhD thesis \cite{Gallager1963}, are ideal for fast communication as they offer lower error floors and higher coding gains at lower signal-to-noise ratio (SNR) than turbo codes \cite{Richardson2018, Mrinmayi2020}. In particular, quasi-cyclic LDPC codes are the standard set by the Third Generation Partnership Project (3GPP) for rate matching and hybrid automatic repeat request (HARQ) \cite{Fossorier2004,Yahya2009,Chiu2018,release18}.  Other applications include storage in solid-state drives \cite{Zhao2013}; and DVB-S2, a satellite-based digital video broadcast standard upgrading the use of Reed-Solomon codes in its previous iteration, DVB-S \cite{dvbs22009}.

The belief propagation algorithm is currently the industry-standard method for decoding LDPC codes \cite{Pearl1982}. It is a message-passing algorithm that performs inference on the Tanner graph of the parity-check matrix of the code. Even though, in general, message-passing algorithms are the most efficient LDPC decoding method due to the low density of LDPC codes \cite{MacKay1999}, they provide exact solutions only for tree graphs \cite{Gallager1963}. For general graphs, message-passing algorithms can only approximate the probability of a bit being in error. Furthermore, despite performing well in practice, the algorithm is not guaranteed to converge in the presence of cycles \cite{Ihler2005}. Cycles, commonly present in short codes, prevent the belief propagation algorithm from reaching maximum likelihood performance. These considerations make belief propagation a suboptimal algorithm, inspiring research to improve its accuracy and expedite communication \cite{Chen2005,Wen2009,Jing2014}. In particular, modifications of belief propagation have been proposed to reduce the error-correcting gap with maximum likelihood with a focus on short codes \cite{Zhang2023}.

Decoding an LDPC code, as with any linear code, involves retrieving original information from received messages despite potential errors introduced during transmission. This process presents a combinatorial optimization challenge: identifying the most likely transmitted codeword (encoded information without error) among all possible codewords based on the received signal. Introduced by Farhi \textit{et~al.}\ in 2014 \cite{Farhi2014}, QAOA is a parameterized quantum optimization algorithm designed to give approximate solutions to combinatorial optimization problems. QAOA seeks the qubit configuration that minimizes a given cost function by iteratively adjusting a set of parameters. QAOA's suitability for solving optimization problems makes it an attractive method for decoding classical linear codes. Notably, three main approaches using QAOA have been proposed: minimum-distance decoding \cite{Matsumine2019}, syndrome decoding \cite{Lai2022}, and minimum distance decoding as a quadratic unconstrained binary optimization (QUBO) problem \cite{kasi2024}.

In this paper, building on previous work by Lai \textit{et~al}.\ on syndrome decoding with QAOA \cite{Lai2022}, we introduce the quantum-enhanced belief propagation algorithm (QEBP), which treats QAOA as a pre-processing step to the belief propagation algorithm. We show through exact simulations that QEBP lowers the average block error rate (BLER) when compared with QAOA syndrome decoding and belief propagation for codes with block length 12. It also reduces the iterations needed for the convergence of belief propagation by $35\%$ on average. We also investigate the decoding of the repetition code with QAOA, as its simplicity allows us to study the scaling of the BLER and the possibility of having shared parameters valid for different decoding rounds. We present a method based on transfer matrices, which allows us to optimize the variational parameters to maximize the probability of successful decoding for any decoding instance of any size. Even though suboptimal, these parameters will enable us to compare the scaling of the BLER for different post-processing strategies.

The paper is organized as follows. In \cref{section:work}, we discuss previous work on quantum approaches to decoding linear codes. Then, in \cref{section:background}, we present LDPC codes, belief propagation decoding, and an overview of QAOA and QAOA syndrome decoding. In \cref{section:bpqaoa}, we describe the QEBP algorithm and, through exact simulations, analyze its error performance and convergence. Finally, in \cref{section:decode_protocol}, we evaluate different QAOA post-processing techniques for decoding in the case of the repetition code.

\section{Related work} \label{section:work}
Bruck and Blaum introduced the connection between error-correcting codes and energy functions in 1989  \cite{Bruck1989}. However, it wasn't until 2019 that the first work on decoding linear codes with QAOA was realized by Matsumine \textit{et al.} describing how to perform minimum distance decoding by defining a cost Hamiltonian, which rewards messages close to the desired output \cite{Matsumine2019}. In their work, they provide an analytic expression of the cost function for decoding Hamming codes with a single QAOA circuit layer. Using this expression, they obtain the minimum cost expectation for $[7,4]$ Hamming codes of different average column degrees, concluding that the lower the average degree distribution, the lower the minimum cost expectation. Note that, in this work, they define the optimization part as maximizing the objective function instead of minimizing it. This indicated that codes with low-density generator matrices are well-suited for level-1 QAOA decoding. However, in practical applications, LDPC codes achieve the highest rates with the least complexity. Unfortunately, LDPC codes do not generally produce low-density generator matrices.

With this motivation in mind, Lai \textit{et al}.\ proposed a method using QAOA analogous to syndrome decoding, defined on the parity-check matrix. In this case, the cost function has a parity-check satisfaction term and an error-weight minimization term \cite{Lai2022}. This work presents decoding performance results for the $[7,4]$ Hamming code and an equivalent code generated through the cyclic permutation of a row vector. Through simulations, they could show near maximum-likelihood block error rates for decoding the Hamming code with four QAOA layers.

To aid with the parameter optimization required for QAOA, Zhu \textit{et~al}.\ employed a meta-learning strategy to train Recurrent Neural Networks \cite{Zhu2024}, which provides initial parameters for the QAOA algorithm. This method improved the bit error rate compared with a random initialization for an LDPC code of size 12 and one layer of QAOA.

Preceding the application of QAOA to decode LDPC codes, considerable work has been done on decoding with quantum annealers \cite{Bian2014,Chancellor2016,Naoki2020}. Of particular note is work on LDPC decoding using a QUBO formulation with a quantum annealer \cite{Kasi2020}. This approach defines a cost function with two terms: a parity-check satisfaction term and a term that minimizes the distance (proportionally to the error parameter of the channel) to the desired codeword. This approach was then adapted to QAOA with the additional observation that the QAOA variational parameters can be reused across decoding rounds with the same signal-to-noise ratio, thus removing the need for parameter optimization at decoding time \cite{kasi2024}. The method presented in this paper shows that the block error rate closely approaches that of belief propagation decoding for a length-13 LDPC code but doesn't give better error correction performance.

On a similar line of thought to our quantum-enhanced decoding method, Dupont \textit{et~al}.\ used QAOA to enhance the relax-and-round algorithm \cite{Dupont2024}. The authors experimentally realize the algorithm using a superconducting quantum computer, achieving similar performance to the Gurobi solver for MaxCut but short of simulated annealing and the Burer-Monteiro algorithm.

%BPQM: \cite{Renes2017,Piveteau2022}
\section{Background} \label{section:background}

In this section, we introduce LDPC codes, a linear code characterized by its sparse parity-check matrix. Then, we give a general description of the belief propagation algorithm, highlighting the key characteristics that affect its performance. Finally, we briefly describe QAOA and the QAOA syndrome decoding algorithm.

%------------------------------------------------------------------
\subsection{Low-density parity-check codes}
Linear codes can be defined for non-binary alphabets, but, in this work, we shall consider only binary linear codes. A linear code $\mathcal{C}$ of length $n$ and dimension $k$ is a $k$-dimensional subspace of the vector space $\mathbb{F}_2^n$ of binary tuples of length $n$ over the field $\mathbb{F}_2$. The vectors $\boldsymbol{x}$ in the codespace $\mathbb{F}_2^n$ are called \textit{codewords}. In general, a code of length $n$ and dimension $k$ is referred to as an $[n,k]$ linear code. The \textit{Hamming distance} between two vectors $\boldsymbol{v}$ and $\boldsymbol{u}$, $d_H(\boldsymbol{v},\boldsymbol{u})$, is the number of positions at which the corresponding bits are different. The \textit{Hamming weight} of a vector $\boldsymbol{v}$, $w_H(\boldsymbol{v})$, is the number of nonzero entries in the vector.

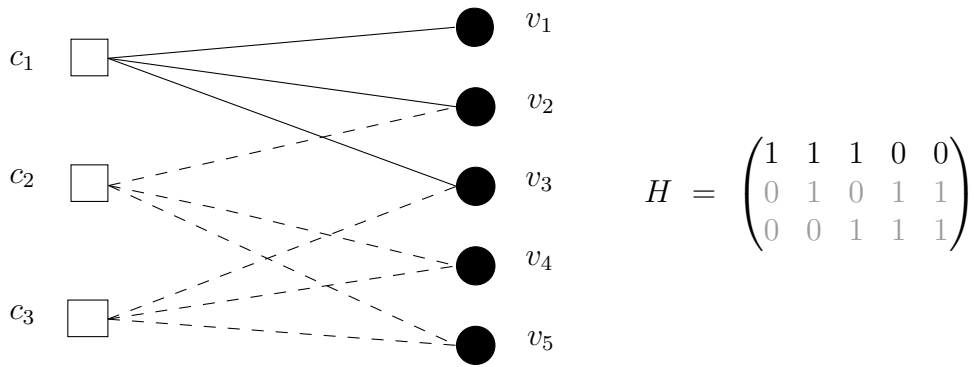
\begin{figure}[ht]
    \centering
    \begin{tikzpicture}[x=0.6pt,y=0.6pt,yscale=-1,xscale=1]
    %uncomment if require: \path (0,288); %set diagram left start at 0, and has height of 288
    \tikzset{every picture/.style={line width=0.75pt}} %set default line width to 0.75pt  
    %Shape: Square [id:dp11307208106305855] 
    \draw  [color={rgb, 255:red, 0; green, 0; blue, 0 }  ,draw opacity=1 ] (91,53) -- (114,53) -- (114,76) -- (91,76) -- cycle ;
    %Shape: Circle [id:dp8503782577020205] 
    \draw  [fill={rgb, 255:red, 0; green, 0; blue, 0 }  ,fill opacity=1 ] (331.5,45.5) .. controls (331.5,38.87) and (335.87,33.5) .. (342.5,33.5) .. controls (350.13,33.5) and (354.5,38.87) .. (354.5,45.5) .. controls (354.5,52.13) and (350.13,57.5) .. (342.5,57.5) .. controls (335.87,57.5) and (331.5,52.13) .. (331.5,45.5) -- cycle ;
    %Shape: Circle [id:dp3367699434937832] 
    \draw  [fill={rgb, 255:red, 0; green, 0; blue, 0 }  ,fill opacity=1 ] (331.5,245.5) .. controls (331.5,238.87) and (336.87,233.5) .. (343.5,233.5) .. controls (350.13,233.5) and (355.5,238.87) .. (355.5,245.5) .. controls (355.5,252.13) and (350.13,257.5) .. (343.5,257.5) .. controls (336.87,257.5) and (331.5,252.13) .. (331.5,245.5) -- cycle ;
    %Shape: Circle [id:dp31810966319496536] 
    \draw  [fill={rgb, 255:red, 0; green, 0; blue, 0 }  ,fill opacity=1 ] (331.5,145.5) .. controls (331.5,138.87) and (336.87,133.5) .. (343.5,133.5) .. controls (350.13,133.5) and (355.5,138.87) .. (355.5,145.5) .. controls (355.5,152.13) and (350.13,157.5) .. (343.5,157.5) .. controls (336.87,157.5) and (331.5,152.13) .. (331.5,145.5) -- cycle ;
    %Shape: Circle [id:dp9101984418738094] 
    \draw  [fill={rgb, 255:red, 0; green, 0; blue, 0 }  ,fill opacity=1 ] (331.5,95.5) .. controls (331.5,88.87) and (336.87,83.5) .. (343.5,83.5) .. controls (350.13,83.5) and (355.5,88.87) .. (355.5,95.5) .. controls (355.5,102.13) and (350.13,107.5) .. (343.5,107.5) .. controls (336.87,107.5) and (331.5,102.13) .. (331.5,95.5) -- cycle ;
    %Shape: Circle [id:dp7697295534340161] 
    \draw  [fill={rgb, 255:red, 0; green, 0; blue, 0 }  ,fill opacity=1 ] (331.5,195.5) .. controls (331.5,188.87) and (336.87,183.5) .. (343.5,183.5) .. controls (350.13,183.5) and (355.5,188.87) .. (355.5,195.5) .. controls (355.5,202.13) and (350.13,207.5) .. (343.5,207.5) .. controls (336.87,207.5) and (331.5,202.13) .. (331.5,195.5) -- cycle ;
    %Shape: Square [id:dp8839854962479061] 
    \draw  [color={rgb, 255:red, 0; green, 0; blue, 0 }  ,draw opacity=1 ] (89,217) -- (114,217) -- (114,240) -- (89,240) -- cycle ;
    %Shape: Square [id:dp9085598959778975] 
    \draw  [color={rgb, 255:red, 0; green, 0; blue, 0 }  ,draw opacity=1 ] (91,132) -- (114,132) -- (114,155) -- (91,155) -- cycle ;
    %Straight Lines [id:da6546117954208617] 
    \draw    (114,65) -- (330.5,45.5) ;
    %Straight Lines [id:da828243583320941] 
    \draw    (114,65) -- (331.5,95.5) ;
    %Straight Lines [id:da7782975084517045] 
    \draw    (114,65) -- (331.5,145.5) ;
    %Straight Lines [id:da6299356923570818] 
    \draw  [dash pattern={on 4.5pt off 4.5pt}]  (114.5,145) -- (331.5,95.5) ;
    %Straight Lines [id:da5042325037266953] 
    \draw  [dash pattern={on 4.5pt off 4.5pt}]  (114.5,145) -- (186.93,160.33) -- (331.5,195.5) ;
    %Straight Lines [id:da4164481511295297] 
    \draw  [dash pattern={on 4.5pt off 4.5pt}]  (114.5,145) -- (331.5,245.5) ;
    %Straight Lines [id:da7204825031704294] 
    \draw  [dash pattern={on 4.5pt off 4.5pt}]  (114,229) -- (331.5,145.5) ;
    %Straight Lines [id:da5458279452829327] 
    \draw  [dash pattern={on 4.5pt off 4.5pt}]  (114,229) -- (331.5,195.5) ;
    %Straight Lines [id:da7569436962039466] 
    \draw  [dash pattern={on 4.5pt off 4.5pt}]  (114,229) -- (331.5,245.5) ;
    
    % Text Node
    \draw (51,59) node [anchor=north west][inner sep=0.75pt]  [xscale=1,yscale=1] [align=left] {$\displaystyle c_{1}$};
    % Text Node
    \draw (51,219) node [anchor=north west][inner sep=0.75pt]  [xscale=1,yscale=1] [align=left] {$\displaystyle c_{3}$};
    % Text Node
    \draw (51,133) node [anchor=north west][inner sep=0.75pt]  [xscale=1,yscale=1] [align=left] {$\displaystyle c_{2}$};
    % Text Node
    \draw (373,35) node [anchor=north west][inner sep=0.75pt]  [xscale=1,yscale=1] [align=left] {$\displaystyle v_{1}$};
    % Text Node
    \draw (374,235) node [anchor=north west][inner sep=0.75pt]  [xscale=1,yscale=1] [align=left] {$\displaystyle v_{5}$};
    % Text Node
    \draw (373,185) node [anchor=north west][inner sep=0.75pt]  [xscale=1,yscale=1] [align=left] {$\displaystyle v_{4}$};
    % Text Node
    \draw (373,135) node [anchor=north west][inner sep=0.75pt]  [xscale=1,yscale=1] [align=left] {$\displaystyle v_{3}$};
    % Text Node
    \draw (374,86) node [anchor=north west][inner sep=0.75pt]  [xscale=1,yscale=1] [align=left] {$\displaystyle v_{2}$};
    % Text Node
    \draw (447,112.4) node [anchor=north west][inner sep=0.75pt]  [xscale=1,yscale=1]  {$H\ =\ \begin{pmatrix}
    1 & 1 & 1 & 0 & 0\\
    \textcolor[rgb]{0.61,0.61,0.61}{0} & \textcolor[rgb]{0.61,0.61,0.61}{1} & \textcolor[rgb]{0.61,0.61,0.61}{0} & \textcolor[rgb]{0.61,0.61,0.61}{1} & \textcolor[rgb]{0.61,0.61,0.61}{1}\\
    \textcolor[rgb]{0.61,0.61,0.61}{0} & \textcolor[rgb]{0.61,0.61,0.61}{0} & \textcolor[rgb]{0.61,0.61,0.61}{1} & \textcolor[rgb]{0.61,0.61,0.61}{1} & \textcolor[rgb]{0.61,0.61,0.61}{1}
    \end{pmatrix}$};

\end{tikzpicture}
    \caption{Tanner graph representation of a linear code. On the left, the Tanner graph with check nodes $c_i$ and variable nodes $v_i$ and, on the right, the corresponding parity-check matrix.}
    \label{fig:tanner_graph}
\end{figure}

Each codeword $\boldsymbol{x} \in \mathcal{C}$ can be written in terms of the basis that spans $\mathcal{C}$. In particular, 
the spanning matrix $G$ for a linear code $\mathcal{C}$ is a $(k \times n)$ matrix such that the rows of $G$ are the basis vectors spanning $\mathcal{C}$ and $\boldsymbol{x} = \boldsymbol{u}G$, where $\boldsymbol{u}$ is the original message of length $k$. This matrix is given the name \textit{generator matrix}. The $(n-k)$-dimensional null space $\mathcal{C}^\bot$ of $G$ consists of all vectors $\boldsymbol{x} \in \mathbb{F}_2^n$ for which $\boldsymbol{x}G^T = \boldsymbol{0}$. The spanning matrix for $\mathcal{C}^\bot$ is the so-called $(n-k)\times n$ \textit{parity-check matrix}, H. This parity check matrix has $(n-k)$ linearly independent rows, each corresponding to a parity-check equation, which all codewords must satisfy. It can be represented as a factor graph, or Tanner graph, where $k$ factors, or check nodes, are connected to the variable nodes that take part in their respective parity-check equation \cite{Tanner1981}, as can be seen in \cref{fig:tanner_graph}. This representation is the basis for message passing or iterative decoding algorithms, such as belief propagation.

A $(d_v, d_c)$-regular LDPC code is a binary linear code in which every codeword bit participates in exactly $d_v$ parity-check equations, and every check equation involves $d_c$ bits. They are \textit{parity-check} codes as they are generated through their parity-check matrix and are \textit{low-density} as the number of ones in the parity-check matrix is lower than the number of zeros. In fact, the fraction of nonzero entries grows linearly with the block length $n$. In contrast, the number of nonzero entries in the parity check matrix of a random linear code is expected to grow as $n^2$ \cite{Richardson2001}.

% For any code $\mathcal{C}$ in $\mathbb{F}^n$, the \textit{dual code} of $\mathcal{C}$, represented by $C^{\bot}$, is defined as
% \begin{equation}
%     \mathcal{C}^\bot = \{ \boldsymbol{v} \in \mathbb{F}^n | \boldsymbol{v} \cdot \boldsymbol{x} = 0, \forall \boldsymbol{x} \in \mathcal{C} \},
% \end{equation}
% where the dot represents the usual dot product and $\boldsymbol{x}$ are the codewords belonging to $\mathcal{C}$.

% \am{Should we define what ``low density'' means in this section?}

%------------------------------------------------------------------
\subsection{Communication channels} \label{section:commchannels}

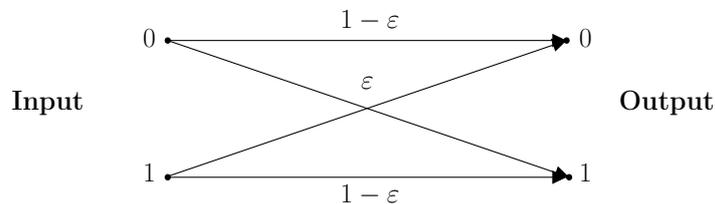
\begin{figure}
    \centering
    % \tikzset{every picture/.style={line width=0.75pt}} %set default line width to 0.75pt        

\begin{tikzpicture}[x=0.6pt,y=0.6pt,yscale=-1,xscale=1]
%uncomment if require: \path (0,258); %set diagram left start at 0, and has height of 258

%Straight Lines [id:da8407107210227749] 
\draw    (205.26,86.34) -- (451.14,86.34) ;
\draw [shift={(454.14,86.34)}, rotate = 180] [fill={rgb, 255:red, 0; green, 0; blue, 0 }  ][line width=0.08]  [draw opacity=0] (8.93,-4.29) -- (0,0) -- (8.93,4.29) -- cycle    ;
%Straight Lines [id:da6088247549079069] 
\draw    (205.26,172.33) -- (397.5,172.33) -- (451.14,172.33) ;
\draw [shift={(454.14,172.33)}, rotate = 180] [fill={rgb, 255:red, 0; green, 0; blue, 0 }  ][line width=0.08]  [draw opacity=0] (8.93,-4.29) -- (0,0) -- (8.93,4.29) -- cycle    ;
%Shape: Ellipse [id:dp6094523448957059] 
\draw  [fill={rgb, 255:red, 0; green, 0; blue, 0 }  ,fill opacity=1 ] (203.67,86.34) .. controls (203.67,85.42) and (204.38,84.67) .. (205.26,84.67) .. controls (206.15,84.67) and (206.86,85.42) .. (206.86,86.34) .. controls (206.86,87.26) and (206.15,88.01) .. (205.26,88.01) .. controls (204.38,88.01) and (203.67,87.26) .. (203.67,86.34) -- cycle ;
%Shape: Ellipse [id:dp35945658192891505] 
\draw  [fill={rgb, 255:red, 0; green, 0; blue, 0 }  ,fill opacity=1 ] (454.14,172.33) .. controls (454.14,171.4) and (454.85,170.66) .. (455.74,170.66) .. controls (456.62,170.66) and (457.33,171.4) .. (457.33,172.33) .. controls (457.33,173.25) and (456.62,174) .. (455.74,174) .. controls (454.85,174) and (454.14,173.25) .. (454.14,172.33) -- cycle ;
%Shape: Ellipse [id:dp8607930077064851] 
\draw  [fill={rgb, 255:red, 0; green, 0; blue, 0 }  ,fill opacity=1 ] (452.54,86.34) .. controls (452.54,85.42) and (453.25,84.67) .. (454.14,84.67) .. controls (455.02,84.67) and (455.74,85.42) .. (455.74,86.34) .. controls (455.74,87.26) and (455.02,88.01) .. (454.14,88.01) .. controls (453.25,88.01) and (452.54,87.26) .. (452.54,86.34) -- cycle ;
%Shape: Ellipse [id:dp4519951287293611] 
\draw  [fill={rgb, 255:red, 0; green, 0; blue, 0 }  ,fill opacity=1 ] (203.67,172.33) .. controls (203.67,171.4) and (204.38,170.66) .. (205.26,170.66) .. controls (206.15,170.66) and (206.86,171.4) .. (206.86,172.33) .. controls (206.86,173.25) and (206.15,174) .. (205.26,174) .. controls (204.38,174) and (203.67,173.25) .. (203.67,172.33) -- cycle ;
%Straight Lines [id:da2902028437892026] 
\draw    (205.26,86.34) -- (452.9,171.35) ;
\draw [shift={(455.74,172.33)}, rotate = 198.95] [fill={rgb, 255:red, 0; green, 0; blue, 0 }  ][line width=0.08]  [draw opacity=0] (8.93,-4.29) -- (0,0) -- (8.93,4.29) -- cycle    ;
%Straight Lines [id:da5892033200531415] 
\draw    (203.67,172.33) -- (451.3,87.31) ;
\draw [shift={(454.14,86.34)}, rotate = 161.05] [fill={rgb, 255:red, 0; green, 0; blue, 0 }  ][line width=0.08]  [draw opacity=0] (8.93,-4.29) -- (0,0) -- (8.93,4.29) -- cycle    ;

% Text Node
\draw (189,78) node [anchor=north west][inner sep=0.75pt]  [xscale=0.6,yscale=0.6] [align=left] {\Large $\displaystyle  0$};
% Text Node
\draw (461,78) node [anchor=north west][inner sep=0.75pt]  [xscale=0.6,yscale=0.6] [align=left] {\Large $\displaystyle  0$};
% Text Node
\draw (189,163) node [anchor=north west][inner sep=0.75pt]  [xscale=0.6,yscale=0.6] [align=left] {\Large $\displaystyle  1$};
% Text Node
\draw (461,163) node [anchor=north west][inner sep=0.75pt]  [xscale=0.6,yscale=0.6] [align=left] {\Large $\displaystyle  1$};
% Text Node
\draw (325,108) node [anchor=north west][inner sep=0.75pt]  [xscale=0.6,yscale=0.6] [align=left] {\Large $\displaystyle  \varepsilon $};
% Text Node
\draw (312,66) node [anchor=north west][inner sep=0.75pt]  [xscale=0.6,yscale=0.6] [align=left] {\Large $\displaystyle  1-\varepsilon $};
% Text Node
\draw (312,176) node [anchor=north west][inner sep=0.75pt]  [xscale=0.6,yscale=0.6] [align=left] {\Large $\displaystyle  1-\varepsilon $};
% Text Node
\draw (107,117) node [anchor=north west][inner sep=0.75pt]  [xscale=0.6,yscale=0.6] [align=left] {\Large \textbf{Input}};
% Text Node
\draw (486,117) node [anchor=north west][inner sep=0.75pt]  [xscale=0.6,yscale=0.6] [align=left] {\Large \textbf{Output}};

\end{tikzpicture}
    \caption{Binary symmetric channel with crossover probability $\varepsilon$.}
    \label{fig:bsc}
\end{figure}

In this work, we study decoding under transmission through the binary symmetric channel (BSC) and the Additive White Gaussian Noise (AWGN) channel with binary-phase shift key (BPSK) modulation. The BSC is a communications channel model that transmits a single bit at a time. This bit may be transmitted in error, or ``flipped'', with \textit{crossover probability} $\varepsilon$, as shown in \cref{fig:bsc}.

An AWGN channel is a noise model used in communication systems to model the noise experienced by a receiver. The noise is modeled as a Gaussian random process, and it is referred to as ``white'' noise as it has a constant power spectral density across all frequencies. Then, the received signal is given by the addition of the transmitted signal and the noise component. Although the noise in some channels may be non-Gaussian, most communication channels of practical interest experience noise due to thermal fluctuations, and so can be modeled as AWGN \cite{Safak2017}. Binary phase shift keying is a digital modulation scheme whereby the information is encoded in the phase of a carrier signal by assigning each bit value to two distinct phases. Using this modulation scheme, one bit of information corresponds to a single phase state. Here, we consider a carrier signal under BPSK affected by AWGN, which, after hard-decision demodulation, behaves like the BSC (\cref{section:comm_channel}). Hence, for simplicity, we will define the QEBP algorithm in terms of the BSC.

%------------------------------------------------------------------
\subsection{Belief propagation algorithm} \label{section:bp}
Belief propagation \cite{Gallager1963} is a message-passing algorithm that performs inference on a factor graph. In the case of LDPC decoding, it is used to calculate the probability of a particular bit being 0 or 1. Two main processes form the general belief propagation algorithm: message-passing and belief estimation.

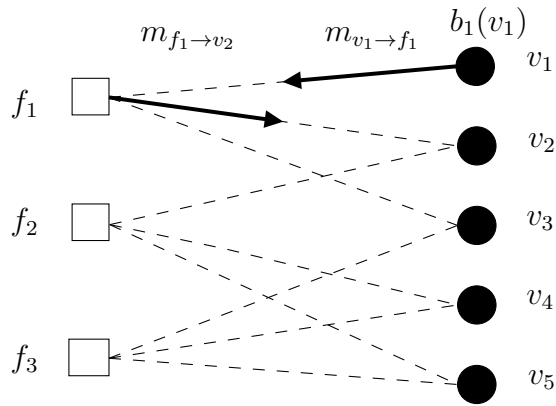
\begin{figure}
    \centering
    \begin{tikzpicture}[x=0.6pt,y=0.6pt,yscale=-1,xscale=1]
%uncomment if require: \path (0,291); %set diagram left start at 0, and has height of 291
\tikzset{every picture/.style={line width=0.75pt}} %set default line width to 0.75pt  
%Shape: Square [id:dp11307208106305855] 
\draw  [color={rgb, 255:red, 0; green, 0; blue, 0 }  ,draw opacity=1 ] (200,59) -- (223,59) -- (223,82) -- (200,82) -- cycle ;
%Shape: Circle [id:dp8503782577020205] 
\draw  [fill={rgb, 255:red, 0; green, 0; blue, 0 }  ,fill opacity=1 ] (439.5,51.5) .. controls (439.5,44.87) and (444.87,39.5) .. (452.5,39.5) .. controls (458.13,39.5) and (463.5,44.87) .. (463.5,51.5) .. controls (463.5,58.13) and (458.13,63.5) .. (452.5,63.5) .. controls (444.87,63.5) and (439.5,58.13) .. (439.5,51.5) -- cycle ;
%Shape: Circle [id:dp3367699434937832] 
\draw  [fill={rgb, 255:red, 0; green, 0; blue, 0 }  ,fill opacity=1 ] (440.5,251.5) .. controls (440.5,244.87) and (445.87,239.5) .. (452.5,239.5) .. controls (459.13,239.5) and (464.5,244.87) .. (464.5,251.5) .. controls (464.5,258.13) and (459.13,263.5) .. (452.5,263.5) .. controls (445.87,263.5) and (440.5,258.13) .. (440.5,251.5) -- cycle ;
%Shape: Circle [id:dp31810966319496536] 
\draw  [fill={rgb, 255:red, 0; green, 0; blue, 0 }  ,fill opacity=1 ] (440.5,151.5) .. controls (440.5,144.87) and (445.87,139.5) .. (452.5,139.5) .. controls (459.13,139.5) and (464.5,144.87) .. (464.5,151.5) .. controls (464.5,158.13) and (459.13,163.5) .. (452.5,163.5) .. controls (445.87,163.5) and (440.5,158.13) .. (440.5,151.5) -- cycle ;
%Shape: Circle [id:dp9101984418738094] 
\draw  [fill={rgb, 255:red, 0; green, 0; blue, 0 }  ,fill opacity=1 ] (440.5,101.5) .. controls (440.5,94.87) and (445.87,89.5) .. (452.5,89.5) .. controls (459.13,89.5) and (464.5,94.87) .. (464.5,101.5) .. controls (464.5,108.13) and (459.13,113.5) .. (452.5,113.5) .. controls (445.87,113.5) and (440.5,108.13) .. (440.5,101.5) -- cycle ;
%Shape: Circle [id:dp7697295534340161] 
\draw  [fill={rgb, 255:red, 0; green, 0; blue, 0 }  ,fill opacity=1 ] (440.5,201.5) .. controls (440.5,194.87) and (445.87,189.5) .. (452.5,189.5) .. controls (459.13,189.5) and (464.5,194.87) .. (464.5,201.5) .. controls (464.5,208.13) and (459.13,213.5) .. (452.5,213.5) .. controls (445.87,213.5) and (440.5,208.13) .. (440.5,201.5) -- cycle ;
%Shape: Square [id:dp8839854962479061] 
\draw  [color={rgb, 255:red, 0; green, 0; blue, 0 }  ,draw opacity=1 ] (198,223) -- (223,223) -- (223,246) -- (198,246) -- cycle ;
%Shape: Square [id:dp9085598959778975] 
\draw  [color={rgb, 255:red, 0; green, 0; blue, 0 }  ,draw opacity=1 ] (200,138) -- (223,138) -- (223,161) -- (200,161) -- cycle ;
%Straight Lines [id:da6546117954208617] 
\draw  [dash pattern={on 4.5pt off 4.5pt}]  (223,71) -- (439.5,51.5) ;
%Straight Lines [id:da828243583320941] 
\draw  [dash pattern={on 4.5pt off 4.5pt}]  (223,71) -- (440.5,101.5) ;
%Straight Lines [id:da7782975084517045] 
\draw  [dash pattern={on 4.5pt off 4.5pt}]  (223,71) -- (440.5,151.5) ;
%Straight Lines [id:da6299356923570818] 
\draw  [dash pattern={on 4.5pt off 4.5pt}]  (223.5,151) -- (440.5,101.5) ;
%Straight Lines [id:da5042325037266953] 
\draw  [dash pattern={on 4.5pt off 4.5pt}]  (223.5,151) -- (295.93,166.33) -- (440.5,201.5) ;
%Straight Lines [id:da4164481511295297] 
\draw  [dash pattern={on 4.5pt off 4.5pt}]  (223.5,151) -- (440.5,251.5) ;
%Straight Lines [id:da7204825031704294] 
\draw  [dash pattern={on 4.5pt off 4.5pt}]  (223,235) -- (440.5,151.5) ;
%Straight Lines [id:da5458279452829327] 
\draw  [dash pattern={on 4.5pt off 4.5pt}]  (223,235) -- (440.5,201.5) ;
%Straight Lines [id:da7569436962039466] 
\draw  [dash pattern={on 4.5pt off 4.5pt}]  (223,235) -- (440.5,251.5) ;
%Straight Lines [id:da6086821174569554] 
\draw [line width=1.5]    (223,71) -- (327.79,85.69) ;
\draw [shift={(331.75,86.25)}, rotate = 187.98] [fill={rgb, 255:red, 0; green, 0; blue, 0 }  ][line width=0.08]  [draw opacity=0] (11.61,-5.58) -- (0,0) -- (11.61,5.58) -- cycle    ;
%Straight Lines [id:da8552442093417947] 
\draw [line width=1.5]    (439.5,51.5) -- (335.23,60.89) ;
\draw [shift={(331.25,61.25)}, rotate = 354.85] [fill={rgb, 255:red, 0; green, 0; blue, 0 }  ][line width=0.08]  [draw opacity=0] (11.61,-5.58) -- (0,0) -- (11.61,5.58) -- cycle    ;

% Text Node
\draw (160,65) node [anchor=north west][inner sep=0.75pt]  [xscale=1,yscale=1] [align=left] {$\displaystyle f_{1}$};
% Text Node
\draw (160,225) node [anchor=north west][inner sep=0.75pt]  [xscale=1,yscale=1] [align=left] {$\displaystyle f_{3}$};
% Text Node
\draw (160,139) node [anchor=north west][inner sep=0.75pt]  [xscale=1,yscale=1] [align=left] {$\displaystyle f_{2}$};
% Text Node
\draw (482,41) node [anchor=north west][inner sep=0.75pt]  [xscale=1,yscale=1] [align=left] {$\displaystyle v_{1}$};
% Text Node
\draw (483,241) node [anchor=north west][inner sep=0.75pt]  [xscale=1,yscale=1] [align=left] {$\displaystyle v_{5}$};
% Text Node
\draw (482,191) node [anchor=north west][inner sep=0.75pt]  [xscale=1,yscale=1] [align=left] {$\displaystyle v_{4}$};
% Text Node
\draw (482,141) node [anchor=north west][inner sep=0.75pt]  [xscale=1,yscale=1] [align=left] {$\displaystyle v_{3}$};
% Text Node
\draw (483,92) node [anchor=north west][inner sep=0.75pt]  [xscale=1,yscale=1] [align=left] {$\displaystyle v_{2}$};
% Text Node
\draw (241,25.4) node [anchor=north west][inner sep=0.75pt]  [xscale=1,yscale=1]  {$m_{f_{1} \rightarrow v_{2}}$};
% Text Node
\draw (356,25.4) node [anchor=north west][inner sep=0.75pt]  [xscale=1,yscale=1]  {$m_{v_{1} \rightarrow f_{1}}$};
% Text Node
\draw (434,12.4) node [anchor=north west][inner sep=0.75pt]  [xscale=1,yscale=1]  {$b_{1}( v_{1})$};

\end{tikzpicture}
    \caption{Message passing in the belief propagation algorithm. Here, $m_{f_1 \rightarrow v_2}$ is the message from factor $f_1$ to variable node $v_2$, $m_{v_1 \rightarrow f_1}$ is the message from variable node $v_1$ to factor $f_1$ and $b_1(v_1)$ is the belief of variable $v_1$.}
    \label{fig:bp}
\end{figure}

The message-passing stage of the algorithm involves two types of operations, represented in \cref{fig:bp}. Firstly, a factor sends a message to a variable node by marginalizing over the other variables connected to the factor. In this way, the message contains the factor's information about the receiving variable node. Denote the set of factor nodes connected to the $j$th variable by $N(j)$, and the set of variables connected to the $i$th factor node by $M(i)$. The message from a factor to a variable node $m_{f_j \rightarrow v_i}$ has the form
\begin{equation}
    m_{f_j \rightarrow v_i} = \sum_{M(j) \backslash v_i} f_j(M(j)) \prod_{k \in N(j)\backslash i} m_{v_k \rightarrow f_j}. \label{eq:ftov}
\end{equation}
where $f_j(M(j))$ is the so-called factor potential, and $m_{v_k \rightarrow f_j}$ is a message sent from variable node $v_k$ to factor $f_j$. For LDPC decoding, the factor potential corresponds to parity-check satisfaction of the variable nodes involved. The algorithm is initialized by setting the values of $m_{v_k \rightarrow f_j}$, which depend on the received bit string and the error probability of the channel.

By a similar marginalization process, the factors' information is updated from the variable nodes using the information from the other connected factors. In particular, the message from variable to factor node is defined as

\begin{equation}
    m_{v_i \rightarrow f_j} = \prod_{s \in N(i) \backslash j} m_{f_s \rightarrow v_i}. \label{eq:vtof}
\end{equation}

Finally, to finish a full iteration, the variable belief $b_i$ is updated by taking the product of messages incoming from all factors connected to the variable $i$, that is,

\begin{equation}
    b_i(v_i) = \prod_{s \in N(i)} m_{f_s \rightarrow v_i}. \label{eq:belief}
\end{equation}
The belief is then used to determine the corrected string. If the string obtained is a codeword, then the algorithm terminates; otherwise, the message-passing rounds continue until the maximum number of iterations set by the user is reached.

% The first step of the algorithm is to initialize the local belief of each variable node $b_i$ by measuring the signal to noise ratio of the communication channel and calculating the probability of error for a transmitted bit.

% \am{I'm not sure I completely understand the algorithm. It seems that at the end of the iteration, we have updated the variable beliefs $b_i(v_i)$, but for the next iteration, we need to know the check node beliefs $c_j(V_j)$ -- how are these set?}

The belief propagation algorithm is equivalent to maximum-likelihood decoding for tree graphs; but for graphs with cycles, there are no convergence guarantees, even though it performs well in practice \cite{Yedida2004,Ihler2005}. It is particularly suited for LDPC decoding as the complexity of the algorithm grows linearly with the size of the code $n$, as the number of messages passed depends on the number of nonzero entries of the parity-check matrix. However, it is not an optimal decoding strategy in general, and there are a few considerations to take into account when concretizing the decoding procedure. 

Firstly, the number of iterations needed for the algorithm to converge, especially for larger codes, may impact the processing latency, amount of time required to process a LDPC codeword \cite{Hailes2016}. The number of iterations can be set, or there can be an early termination clause that checks if the belief at the end of each iteration results in a codeword. As the code size tends to infinity, assuming the number of iterations is less than or equal to half of the girth, the probability of error in decoding decays exponentially \cite{Richardson2001}. When the number of iterations surpasses this, due to the cycles present in the code, the performance may be hindered. In the case of iterative decoding of LDPC codes under the BSC or AWGN channel, the effect of cycles is studied by defining trapping sets---sets of variables that are never corrected. By identifying trapping sets, one can improve the LDPC code construction, but their removal may affect the density of the code and, therefore, their error-correcting ability \cite{Price2017}.

% \note{rephrase - error exponentially decreases for iterations chosen to be $g/2$} It is known that for a graph with girth $g$, the belief propagation algorithm will stop producing the correct marginal probabilities for all nodes at iteration $g/2$.  \cite{Gallager1963} Nonetheless, in general, the choice of maximum iterations is greater than $g/2$ as the extra iterations, even though they lead to inexact marginal probabilities on the nodes involved in the smallest cycle, improve the beliefs on the other nodes. 

The scheduling scheme for the message passing also influences the processing latency. There are certain LDPC code constructions that allow for parallel decoding \cite{Gomes2007} increasing the processing latency, but fully parallel schemes limit the flexibility of LDPC codes accepted by the decoder. In contrast, a serial decoder allows for full decoding flexibility. As a result, decoding architectures that allow partial parallelization and accept a considerably large family of codes, like quasi-cyclic LDPC codes \cite{Tanner2004,Richardson2018}, have become the industry standard for 5G communications \cite{release18}. 

\subsection{Quantum approximate optimization algorithm}

The Quantum approximate optimization algorithm (QAOA) \cite{Farhi2014} is a variational quantum algorithm that produces approximate solutions for combinatorial optimization problems. In particular, for a classical cost function $C(\boldsymbol{z})$ defined on $n$-bit strings $\boldsymbol{z} \in \{0,1\}^n$, the aim is to find the bit string $\boldsymbol{z}$ that minimizes the cost function. To construct the ansatz for the variational circuit, we define the corresponding cost operator $\widehat{C}$, diagonal in the computational basis, such that 
\begin{equation}
    \widehat{C} \ket{\boldsymbol{z}} = C(\boldsymbol{z}) \ket{\boldsymbol{z}},
\end{equation}
and introduce the following unitary operator, $U(\widehat{C},\gamma)$, depending on said cost operator $\widehat{C}$ and the variational parameter $\gamma$
\begin{equation}
    U(\widehat{C},\gamma) =  e^{-\mathrm{i}\gamma \widehat{C}}. \label{eq:costuni}
\end{equation}
We also introduce the operator $B = \sum_{j=1}^n X_j$, which we refer to as the mixer Hamiltonian. Then, define a unitary operator, $U(B,\beta)$, that depends on $B$ and the variational parameter $\beta$
\begin{equation}
    U(B,\beta) = e^{- \mathrm{i} \beta B}. \label{eq:mixuni}
\end{equation}

To construct the QAOA circuit, initialize the quantum circuit to the uniform superposition state $\ket{+}^{\otimes n}$ and apply $p$ layers of $U(\widehat{C},\gamma)$ and $U(B,\beta)$ in the following manner 
\begin{equation}
    \ket{\psi_{\gamma, \beta}} = U(B,\beta_p)U(\widehat{C},\gamma_p) \ldots U(B,\beta_1)U(\widehat{C},\gamma_1) \ket{+}^{\otimes n}.\label{eq:ansatz}
\end{equation}
The objective function for a cost function $C$ for the variational algorithm is $\bra{\psi_{\gamma, \beta}} \widehat{C} \ket{\psi_{\gamma, \beta}}$. This function is minimized with respect to the variational parameters ${\boldsymbol{\gamma} = (\gamma_1, \gamma_2, \dots, \gamma_p)}$ and ${\boldsymbol{\beta} = (\beta_1, \beta_2, \dots, \beta_p)}$, where $\gamma_k \in [0,2\pi)$ and $\beta_k \in [0,\pi)$ for $k \in [1,p]$. The final solution is obtained by measuring the quantum circuit in the computational basis to obtain a close-to-optimal bit string $\boldsymbol{z}$.

%------------------------------------------------------------------
\subsection{QAOA syndrome decoding algorithm}

A QAOA-based decoder implements a cost function that is minimized for the desired codeword. There are two main approaches to defining the cost function: a minimum-distance decoder and a syndrome-based decoder. The minimum-distance decoder has a single term that is minimized for qubit combinations that belong to the code defined by its generator matrix \cite{Matsumine2019}. In this work, however, we will use the syndrome-based decoder.

In syndrome decoding, the objective is to find the error that satisfies a particular \textit{syndrome}. Let  $\boldsymbol{y} \in\{0,1\}^{n}$ be the vector received at the end of the communication protocol. For a linear code, this vector can be expressed as $\boldsymbol{y} = \boldsymbol{x} + \boldsymbol{\Tilde{e}}$, where $\boldsymbol{x}$ is the original message and $\Tilde{\boldsymbol{e}}$ is the error vector.  If the received string is such that ${\boldsymbol{y} H^{T} = (\boldsymbol{x} + \boldsymbol{\Tilde{e}})H^{T} \neq \boldsymbol{0}}$, then $\boldsymbol{\Tilde{e}}$ has nonzero entries and $\boldsymbol{y}$ is not a codeword. The vector $\boldsymbol{s} = \boldsymbol{y} H^{T}$ is called the \textit{error syndrome} or \textit{syndrome} of $\boldsymbol{y}$. Given a syndrome vector $\boldsymbol{s} \in\{0,1\}^{r}$, where $r=n-k$, the minimum weight decoding rule is to find the minimum weight error $\boldsymbol{\Tilde{e}}$ that satisfies the syndrome equation, that is,
\begin{equation}
    \underset{\boldsymbol{\Tilde{e}} \in\{0,1\}^{n} \text{ s.t. } \boldsymbol{\Tilde{e}} H^{T}=\boldsymbol{s}}{\arg \min } w_H(\boldsymbol{\Tilde{e}}), \label{eq:syndecoding}
\end{equation}
where $w_H$ is the Hamming weight.

Following this idea, define the following cost Hamiltonian for QAOA with $n$ qubits according to syndrome $s=\left(s_{1}, s_{2}, \ldots, s_{r}\right) \in\{0,1\}^{r}$ as
\begin{equation}
\widehat{C}= -\eta \sum_{j=1}^{r}\left(1-2 s_{j}\right) \prod_{\ell=1}^{n} Z_{\ell}^{[H]_{j, \ell}} - \alpha \sum_{j=1}^{n} Z_{j}, \label{eq:costhparity}
\end{equation}
where $\eta$ and $\alpha$ are balancing parameters, and $\prod_{\ell=1}^{n} Z_{\ell}^{[H]_{j, \ell}}$ is the tensor product corresponding to all nonzero elements in the $j$th row of the parity-check matrix. The rows of $H$ correspond to the constraints, and the syndrome tells us whether the constraint is satisfied ($+1$) or not ($-1$). The second term is a penalty function for higher error weights. 

% As every computational basis state is an eigenstate of $\sum_{j=1}^{n} Z_{j}$, a basis vector of lower weight would have higher energy. 
% ($Z=|0\rangle\langle 0|-| 1\rangle\langle 1|$).

% The corresponding eigenvalue equation would be
% \begin{equation}
%     \prod_{\ell=1}^{n} Z_{\ell}^{[H]_{j, \ell}} \ket{\boldsymbol{\Tilde{e}}} = \left(1-2 \left[\boldsymbol{y} H^{T}\right]_j\right)  \ket{\boldsymbol{\Tilde{e}}},
% \end{equation}
% where $\ket{\boldsymbol{\Tilde{e}}}$ denotes the state corresponding to the error on the message $\boldsymbol{\Tilde{e}} \in \{0,1\}^{n}$.

% The output of QAOA for this method is the error in the message, and the operations required are of order $\mathcal{O}(2n-k)$.

% \SetKwComment{Comment}{/* }{ */}

% \begin{algorithm}
% \caption{QAOA syndrome decoding}\label{alg:two}
% \KwData{$H$, $\boldsymbol{s} \in \{0,1\}^r$, $\alpha$, $\eta$.}
% \KwResult{$\Tilde{\boldsymbol{\Tilde{e}}}$}
% % $y \gets 1$\;
% % $X \gets x$\;
% % $N \gets n$\;
% % \While{$N \neq 0$}{
% %   \eIf{$N$ is even}{
% %     $X \gets X \times X$\;
% %     $N \gets \frac{N}{2}$ \Comment*[r]{This is a comment}
% %   }{\If{$N$ is odd}{
% %       $y \gets y \times X$\;
% %       $N \gets N - 1$\;
% %     }
% %   }
% % }
% \end{algorithm}

\section{Quantum-enhanced belief propagation algorithm} \label{section:bpqaoa}
We now outline the quantum-enhanced approach to belief propagation using the min-sum algorithm in the log-likelihood domain due to its numerical stability. The belief propagation algorithm in the log-likelihood domain follows the same steps as described in \cref{section:bp}, using log-likelihood ratio (LLRs) as the messages being sent between check nodes and variable nodes.

% To concretize the messages passed in the algorithms, we define the following quantities that indicate the marginal likelihood of checks being satisfied. 
Our objective is to compute the log-likelihood ratio 
\begin{equation}
    L(v_i) = \log \left( \frac{\mathbb{P} (v_i = 0| y_i)}{\mathbb{P}(v_i = 1| y_i)} \right), \label{eq:vllr} 
\end{equation}
where $\mathbb{P} (v_i = b| y_i)$ is the probability that $v_i = b$ for $b \in \{0,1\}$ given the value of $y_i \in \{0,1\}$. In this way, the sign of the LLR conveys the value of the corresponding bit, whilst its magnitude indicates how likely this value is. The higher the LLR, the greater certainty one has of the value of the particular bit. 

The expression for messages from check nodes to variable nodes given in \cref{eq:ftov} become
\begin{equation}
    m_{c_j \rightarrow v_i} = L(q_{ij}) = \log \left( \frac{q_{ij} (0)}{q_{ij} (1)} \right), \label{eq:ctovllr}
\end{equation}
where $q_{ij} (b)$ is the probability that check node $v_i = b$, given the value of the $i$th entry of the received message $y_i$ and assuming the parity checks involving $c_i$ are satisfied. Note that, as before, a message sent to $v_i$ is independent of the previous message coming from $v_i$. Similarly, the messages from variable nodes to check nodes in \cref{eq:vtof} have the following form,
\begin{equation}
    m_{v_i \rightarrow c_j} = L(c_{ji}) = \log \left( \frac{c_{ji} (0)}{c_{ji} (1)} \right), \label{eq:vtocllr}
\end{equation}
where $c_{ji}(b)$ is the probability that the parity-check $c_j$ is satisfied given $v_i=b$. These messages are independent of the messages previously sent by node $c_j$. Finally, we denote the belief after a full iteration as defined in \cref{eq:belief} as 
\begin{equation}
    b_i =  L(Q_{i}) = \log \left( \frac{Q_{i} (0)}{Q_{i} (1)} \right), \label{eq:bllr}
\end{equation}
where $Q_i(b)$ is the belief that $v_i=b$.
% The messages from check nodes to variable node are 

% Given a check node $c_{i} = b$, where $b \in \{0,1\}$ is its bit value, $c_{ji}(b)$ is the probability that the check equation $j$ is satisfied. For a variable node $v_i$, the message $q_{ij}(b)$ is the probability that a check node $c_i = b$, assuming that the equations involving $c_i$ are satisfied and dependent on the value of the bit received $y_i$ and the messages from all other check nodes except the check node sending the message to $v_i$. The final decision on the bit value is given by $Q_i(b)$ dependent on all check node messages received by $v_i$.

% Now, we can rewrite the messages in terms of log-likelihood ratios dependent on the probabilities just defined as
% \begin{align*}
%     L(v_i) &= \log \left( \frac{\mathbb{P} (v_i = 0| y_i)}{\mathbb{P}(v_i = 1| y_i)} \right), \\
%     L(c_{ji}) &= \log \left( \frac{c_{ji} (0)}{c_{ji} (1)} \right), \\
%     L(q_{ij}) &= \log \left( \frac{q_{ij} (0)}{q_{ij} (1)} \right), \\
%     L(Q_{i}) &= \log \left( \frac{Q_{i} (0)}{Q_{i} (1)} \right),
% \end{align*}
% where $L(v_i)$ is the initial LLR assigned to each variable node $i$.

\subsection{The algorithm} \label{section:alg}
The key idea behind the quantum-enhanced approach is to use the distribution of samples given by QAOA to determine the probability that QAOA introduces a bit correction. This is defined as
\begin{equation}
    \varepsilon^{\mathrm{QAOA}}_i = \sum_{\{\Tilde{\boldsymbol{e}} | \Tilde{e}_i = 1\}} 
 \mathbb{P}_{\mathrm{QAOA}}\left(\Tilde{\boldsymbol{e}} |  \boldsymbol{s} = \boldsymbol{y} H^T\right), \label{eq:qaoa_crossover}
\end{equation}
where $i$ refers to the $i$th bit and
\begin{equation}
    \mathbb{P}_{\mathrm{QAOA}}(\Tilde{\boldsymbol{e}}|  \boldsymbol{s} = \boldsymbol{y} H^T, \alpha, \eta, \boldsymbol{\gamma}, \boldsymbol{\beta}) = \abs{\bra{\Tilde{\boldsymbol{e}}} \prod_{\ell=1}^{p} e^{-\mathrm{i}\beta_\ell B}e^{-\mathrm{i}\gamma_\ell \widehat{C}}\ket{+}}^2, \label{eq:prob_error_qaoa}
\end{equation}
where $p$ is the number of QAOA layers, $\widehat{C}$ is as defined in \cref{eq:costhparity}. If one were to decode the received message using this information, then bit $i$ would flip if $\varepsilon^{\mathrm{QAOA}}_i > 1/2$. Hence, we can treat this as a channel that flips a bit if the error probability of that bit is greater than one half. With this information, we construct a channel that consists of a concatenation of the original BSC channel and the QAOA probability of error output. This channel has the following parameter for each bit $i$:
\begin{equation}
    \varepsilon_i = \left(1-\varepsilon^{\mathrm{BSC}}\right)\varepsilon^{\mathrm{QAOA}}_i + \varepsilon^{\mathrm{BSC}}\left(1-\varepsilon^{\mathrm{QAOA}}_i\right).  \label{eq:error_bpqaoa}
\end{equation}
Hence, the initial log-likelihood rate becomes
\begin{equation}
    L(q_{ij}) = L(v_i) =(-1)^{y_i} \log \left(  \frac{1 - \varepsilon_i}{\varepsilon_i}\right). \label{eqn:llr_qaoa}
\end{equation}
Now, we can look at how the probability of error will impact the value of the initial log-likelihood ratio. We assume in the following that $0 < \varepsilon^{\mathrm{BSC}} < 1/2$. Depending on the value of $\varepsilon_i^{\mathrm{QAOA}}$, we can distinguish three cases:
\begin{itemize}
    \item if $\varepsilon_i^{\mathrm{QAOA}} > 1/2$, then from the QAOA probability distribution, we determine there is an error in position $i$,
    \item if $\varepsilon_i^{\mathrm{QAOA}} < 1/2$, then from the QAOA probability distribution, we determine there is not an error in position $i$,
    \item if $\varepsilon_i^{\mathrm{QAOA}} = 1/2$, then there is no decision on whether there is an error in position $i$.
\end{itemize}
For $\varepsilon_i^{\mathrm{QAOA}} > 1/2$, assuming $0 < \varepsilon^{\mathrm{BSC}} < 1/2$, the final probability of error is $\varepsilon_i > 1/2$. So, 
\begin{equation*}
    0 < \frac{1 - \varepsilon_i}{\varepsilon_i} < 1 \Rightarrow \log \left( \frac{1 - \varepsilon_i}{\varepsilon_i} \right) < 0,
\end{equation*}
and \cref{eqn:llr_qaoa} becomes
\begin{equation}
    L(v_i) = (-1)^{y_i + 1} \left| \log \left(  \frac{1 - \varepsilon_i}{\varepsilon_i}\right) \right|. \label{eqn:llr_qaoa_error}
\end{equation}
Notice that, when $y_i=0$, the log-likelihood ratio is negative, signifying that the corresponding codeword bit should be $1$. Conversely, when $y_i=1$, the log-likelihood is positive, indicating that the codeword bit should be $0$. 
Similarly, if $\varepsilon_i^{\mathrm{QAOA}} < 1/2$, then  $\log \left( (1 - \varepsilon_i)/\varepsilon_i \right) > 0$ and
\begin{equation}
    L(v_i) = (-1)^{y_i} \left| \log \left(  \frac{1 - \varepsilon_i}{\varepsilon_i}\right) \right|, \label{eqn:llr_qaoa_no_error}
\end{equation}
which assigns the correct corresponding signs. Finally, if the probability $\varepsilon_i^{\mathrm{QAOA}} = 1/2$, then ${L(v_i) = \log(1) = 0}$. Here, we can see how the initial QAOA input feeds an initial guess. \Cref{eqn:llr_qaoa_error}, when QAOA detects an error, flips the sign of the corresponding input LLR, while \cref{eqn:llr_qaoa_no_error} maintains it. In the case where $1/2 < \varepsilon^{\mathrm{BSC}} < 1$, the LLR signs for cases 1 and 2 are interchanged, which gives the desired behavior.

Then, the steps of the algorithm are as before. The check node update equation in \cref{eq:vtocllr} becomes

\begin{equation}
    L(c_{ji}) = \prod_{i' \in V_{j \backslash i}} \alpha_{i'j} \cdot \phi \left( \sum_{i' \in V_{j \backslash i}} \phi(\beta_{i'j})\right), \label{eq:cjillr}
\end{equation}
where
\begin{align*}
    \alpha_{ij} &= \mathrm{sign} (L(q_{ij})), \\
    \beta_{ij} &= \left|L(q_{ij})) \right|, \\
    \phi(x) &= -\log \left( \tanh \frac{x}{2}  \right) = \log \left( \frac{e^x +1}{e^x -1} \right).
\end{align*}

For the BSC, this can be further simplified since we can see that the term corresponding to the smallest $\beta_{ij}$ in the summation dominates from the shape of $\phi(x)$. Noting that $\phi(x)$ is its own inverse,
\begin{equation*}
    \phi \left( \sum_{i' \in V_{j \backslash i}} \phi(\beta_{i'j})\right) \simeq \phi \left(\phi\left( \min_{i' \in V_{j \backslash i}} \beta_{i'j} \right)\right) = \min_{i' \in V_{j \backslash i}} \beta_{i'j}.
\end{equation*}
So, \cref{eq:cjillr} can be re-written as
\begin{equation}
    L(c_{ji}) = \prod_{i' \in V_{j \backslash i}} \alpha_{i'j} \cdot \min_{i' \in V_{j \backslash i}} \beta_{i'j}. \label{eq:llr_min_sum}
\end{equation}
The variable node update equation in \cref{eq:ctovllr} takes the form 
\begin{equation}
    L(q_{ij}) = L(v_i) + \sum_{j' \in v_{i\backslash j}} L(c_{j'i}). \label{eq:llr_v_node}
\end{equation}
Finally, we calculate the belief for each bit,
\begin{equation}
    L(Q_{i}) = L(v_i) + \sum_{j \in v_{i}} L(c_{ji}).
\end{equation}
We denote $\widehat{\boldsymbol{v}}$ to be the received vector estimated by our algorithm. We assign values to each $\widehat{v}_i$ by
\begin{equation}
    \widehat{v}_i = \begin{cases}
        1 \hspace{1cm} L(Q_{i}) < 0, \\
        0 \hspace{1cm} \text{otherwise}.
    \end{cases}
\end{equation}

After determining the $\widehat{v}$, we calculate the corresponding syndrome $\boldsymbol{s} = \widehat{\boldsymbol{v}}H^{T}$. If $\boldsymbol{s} = 0$ or a set number of maximum iterations is reached, the algorithm terminates; otherwise, it returns to \cref{eq:llr_min_sum}.

\begin{figure}
    \centering
    \scalebox{1.3}{\input{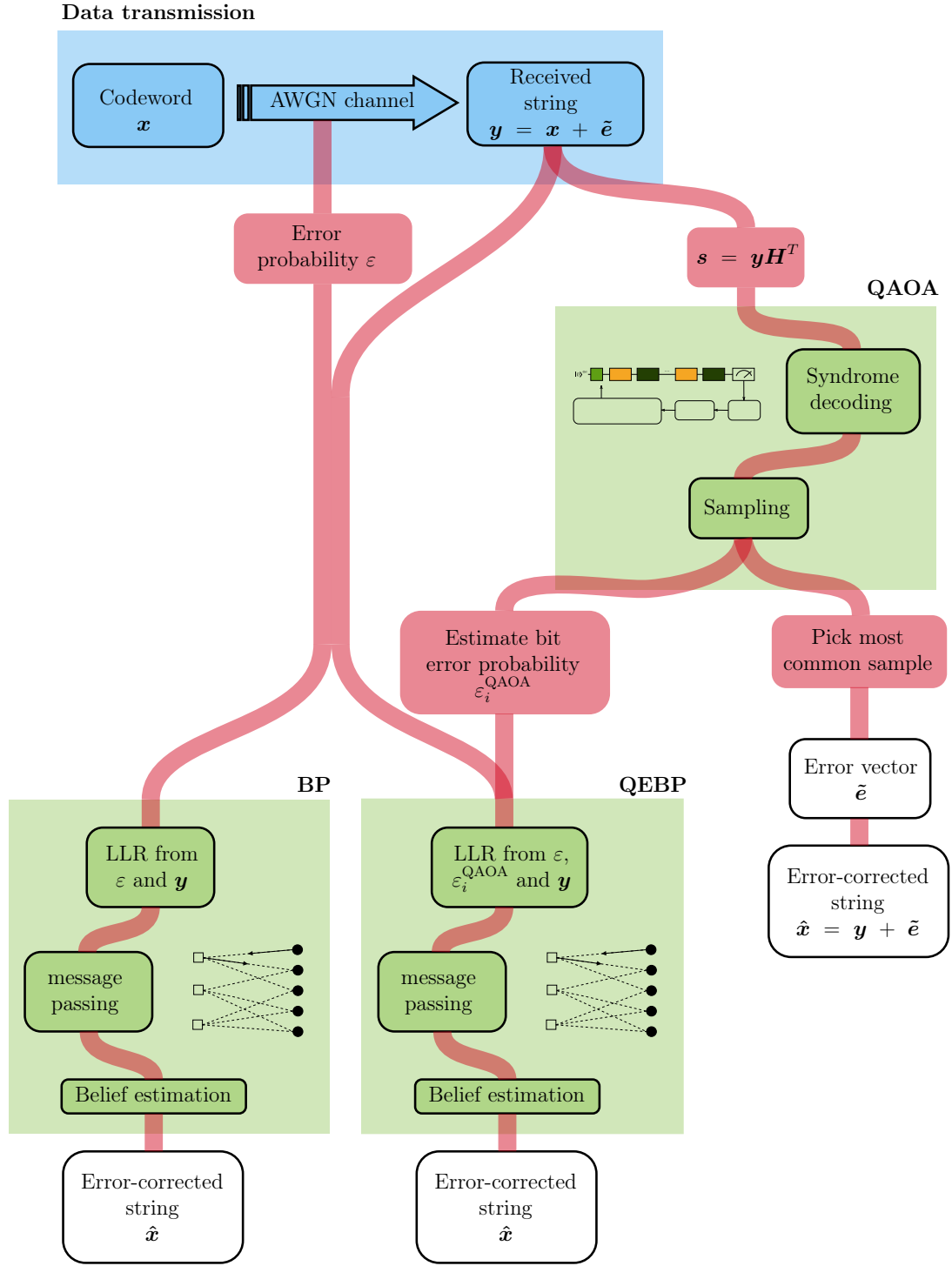}}
    \caption{Decoding process for a single decoding round of belief propagation, quantum-enhanced belief propagation, and syndrome decoding with QAOA.}
    \label{fig:decmeth}
\end{figure}

\Cref{fig:decmeth} shows a single QEBP decoding round, illustrating its connection to standard belief propagation and syndrome decoding with QAOA. As with any communication protocol, data is transmitted through a channel; in this case, the physical noise model we use is the AWGN channel (the connection between the AWGN channel and a BSC channel is explored in \cref{section:comm_channel}), which introduces errors. From this process, we can extract the error probability induced by the channel and find the syndrome associated with the error in the codeword. The QAOA syndrome decoding algorithm only uses the syndrome information without the probability of error, assuming a low-weight error is more probable. In contrast, the belief propagation algorithm takes both the error probability of the channel and the received bit-string as inputs. 

A single round of decoding with the quantum-enhanced belief propagation algorithm involves running the QAOA syndrome decoding algorithm for the syndrome of the string received. Then, from the measurement outcomes, one can estimate the probability of QAOA correcting a bit or, treating QAOA as a second channel, the error probability of the QAOA syndrome decoding algorithm. This estimate, in conjunction with the error probability of the channel and the received string, then becomes the input of the belief propagation algorithm, as previously described. We will consider two cases for the error probability of QAOA. The first, as given in \cref{eq:qaoa_crossover}, calculates the error probability of each bit by computing the marginal probability of each bit being in error. The second, the QEBP one-sample approach, assigns 0 or 1 to the crossover probability depending on the string with the highest number of measurement counts.

\subsection{Equivalence under transmission} \label{section:zerotransmission}
Up to this point, the only assumptions made on the communication protocol are that the codeword belongs to a binary linear code and that all codewords are equally likely to have been transmitted. However, for simplicity in the analysis, one may wish to assume the transmission of the all-zero codeword. To do so, the construction of the LLR input from the BSC channel and QAOA output, as well as the belief propagation algorithm, must satisfy the  following symmetry conditions  \cite{Richardson2001}.

\begin{enumerate}
    \item Channel symmetry. The channel is output symmetric: $${\mathbb{P}(y_i = 1 | x_i = 1) = \mathbb{P}(y_i = 0 | x_i =0)}$$ and $$\mathbb{P}(y_i = 0 | x_i = 1) = \mathbb{P}(y_i = 1 | x_i =0).$$
    
    \item Check node symmetry. Signs factor out of check node message maps, that is,
    \begin{equation*}
        L(c_{ji}) = \prod_{i' \in V_{j \backslash i}} \alpha_{i'j} \cdot \min_{i' \in V_{j \backslash i}} \beta_{i'j}, \label{eq:sym2}
    \end{equation*}
    where $\alpha_{ij} = \mathrm{sign} (L(q_{ij}))$ and $
    \beta_{ij} = \left|L(q_{ij})) \right|$, as before.
    \item Variable node symmetry. Sign inversion invariance of variable node message map holds. At initialization, $-L(v_i) = L(v_i)$. Also, if the signs of the incoming check node messages are reversed, the resulting variable node update expression is $L'(q_{ij}) = -L(q_{ij})$, where $L(q_{ij})$ is the original variable node LLR.
\end{enumerate}

The second and third conditions are assumptions built into the belief propagation algorithm, so the quantum-enhanced approach satisfies them. On the other hand, the first is a condition on the channel. As illustrated in \cref{section:alg}, from the point of view of the belief propagation algorithm, the input is a log-likelihood depending on the BSC channel error probability and the estimated probability of error given by QAOA. \Cref{eq:error_bpqaoa} is obtained by treating the QAOA decoding step as a second BSC channel and concatenating it with the actual BSC channel. Then, using the monotonicity of BSC channels, the combined channels are treated as a single BSC channel with probability of error $\varepsilon_i$ for each bit $i$. For this to be a valid assumption, $\varepsilon_i^{\mathrm{QAOA}}$ must be independent of the codeword transmitted and, so, have a symmetric bit crossover probability. As seen from \cref{eq:prob_error_qaoa}, the probability of obtaining any error string after QAOA decoding only depends on the syndrome, which is itself independent of the original codeword transmitted. Therefore, we can treat the input of the belief propagation algorithm as a single BSC with probability of error $\varepsilon_i$, which satisfies the first condition. 

To show that the output of the quantum-enhanced belief propagation algorithm is invariant under codewords transmission, we invoke the following lemma on the probability of error \cite{Richardson2001}.

\begin{lemma}[Independence of Error Probability Under Symmetry]\label{lemma:allzero}
Let $G$ be a bipartite graph representing a binary linear code, and for a given message-passing algorithm, let $\mathbb{P}_e^{(\ell)}(\boldsymbol{x})$ denote the conditional probability of error after $\ell\mathrm{th}$ decoding iteration, assuming that codeword $\boldsymbol{x}$ was sent. If the channel and the decoder fulfill the symmetry conditions, then $\mathbb{P}_e^{(\ell)}(\boldsymbol{x})$ is independent of $\boldsymbol{x}$.
\end{lemma}

\Cref{lemma:allzero} implies that, by construction, the algorithm does not depend on the original codeword being transmitted. 
% Thus, for all simulations discussed from this point, we will assume the transmission of the all-zero codeword.

\subsection{Error rate performance}

We compare the BLER performance of QAOA-based decoding methods against the min-sum belief propagation algorithm and maximum-likelihood decoding. The methods we concentrate on are the QAOA syndrome decoding algorithm, the QAOA syndrome decoding algorithm with an additional codeword post-selection step, the QEBP algorithm as described in \cref{section:alg}, and QEBP with one sample as input. For the case where a single sample is taken as input, the sample chosen determines the QAOA bit correction probability; that is, for a sample $\boldsymbol{\Tilde{e}}$, if $\Tilde{e}_i = b$, where $b \in \{0,1\}$, then $\varepsilon_i^{\mathrm{QAOA}} = b$. We choose the balancing parameters to be $\alpha = 1$ and $\eta = 2$ to give the parity-check term in the cost Hamiltonian a higher weight. In all cases, we average over 10000 random decoding instances where a codeword has been transmitted through an AWGN channel using BPSK modulation as discussed in Appendix \ref{section:comm_channel}. We show results for two $[12,8]$ LDPC codes, a $(2,3)$-regular code with redundant parity checks, and an irregular code with an average variable node degree of 1.92 and an average check node degree of 2.88. 

\begin{figure}[ht]
\begin{subfigure}{.4226\linewidth}
    \centering
      \includegraphics[width=\linewidth]{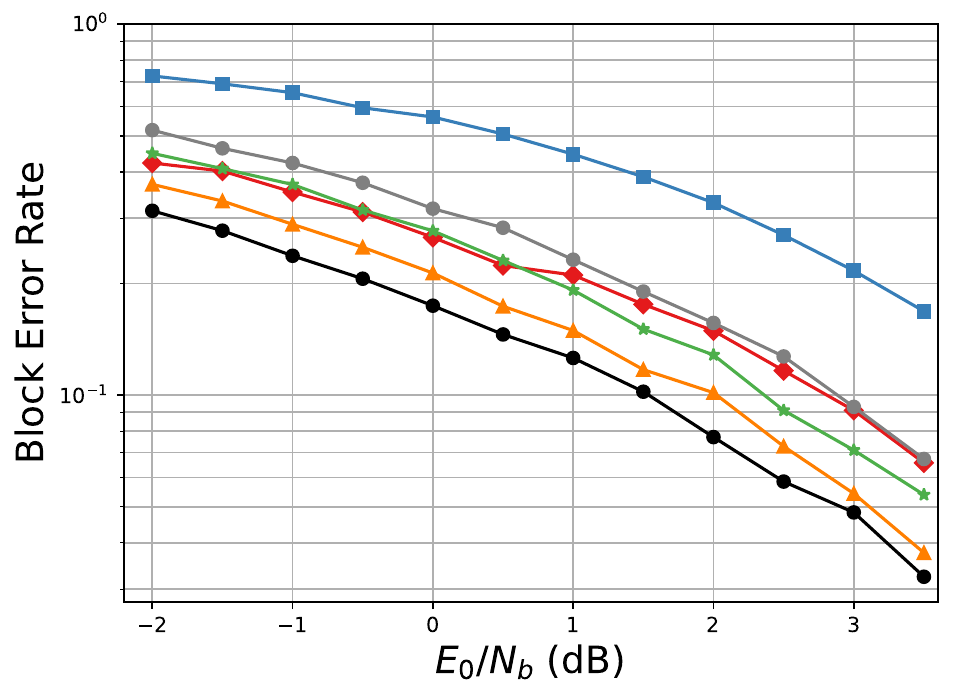}
      \caption{$[12,8]$-regular code.}
      \label{fig:1283r}
\end{subfigure}
\begin{subfigure}{.5774\linewidth}
\centering
  \includegraphics[width=\linewidth]{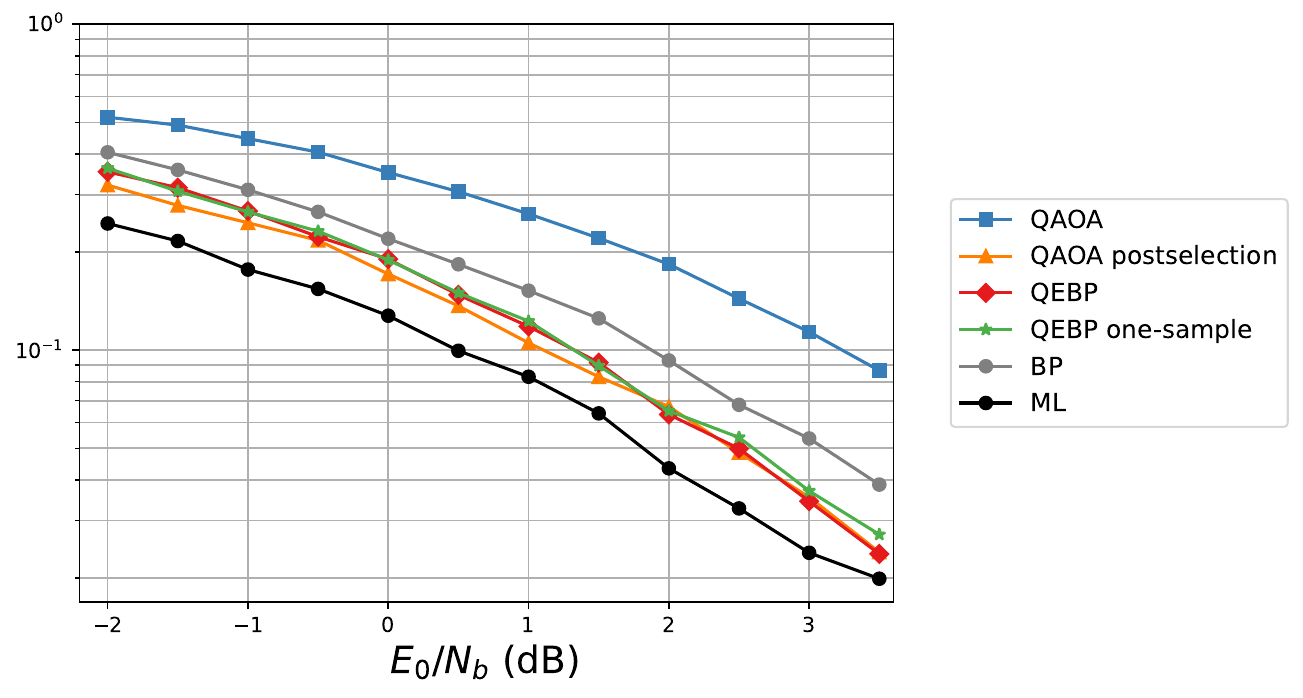}
  \caption{$[12,8]$-irregular code.}
  \label{fig:1283irr}
\end{subfigure}
\caption{Block error rate against signal-to-noise ratio $(E_0/N_b)$ for syndrome decoding with QAOA with $p=3$. We compare results using QAOA syndrome decoding, QAOA with codeword post-selection, and quantum-enhanced belief propagation decoding.}
\label{fig:1283}
\end{figure}

\begin{figure}[ht]
\begin{subfigure}{.4226\linewidth}
    \centering
      \includegraphics[width=\linewidth]{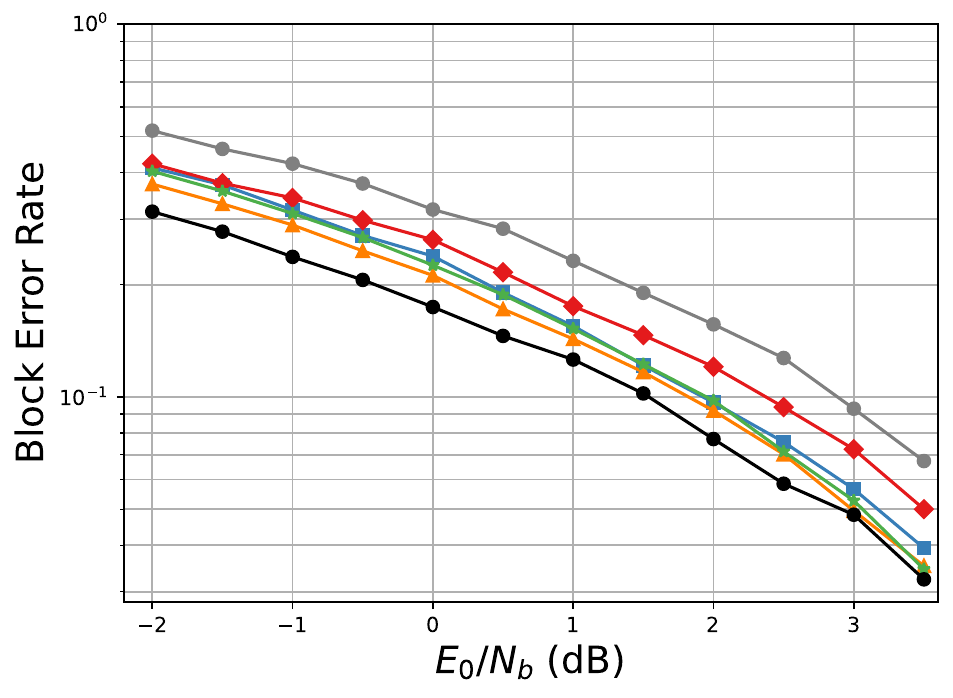}
      \caption{$[12,8]$-regular code.}
      \label{fig:1285r}
\end{subfigure}
\begin{subfigure}{.5774\linewidth}
\centering
  \includegraphics[width=\linewidth]{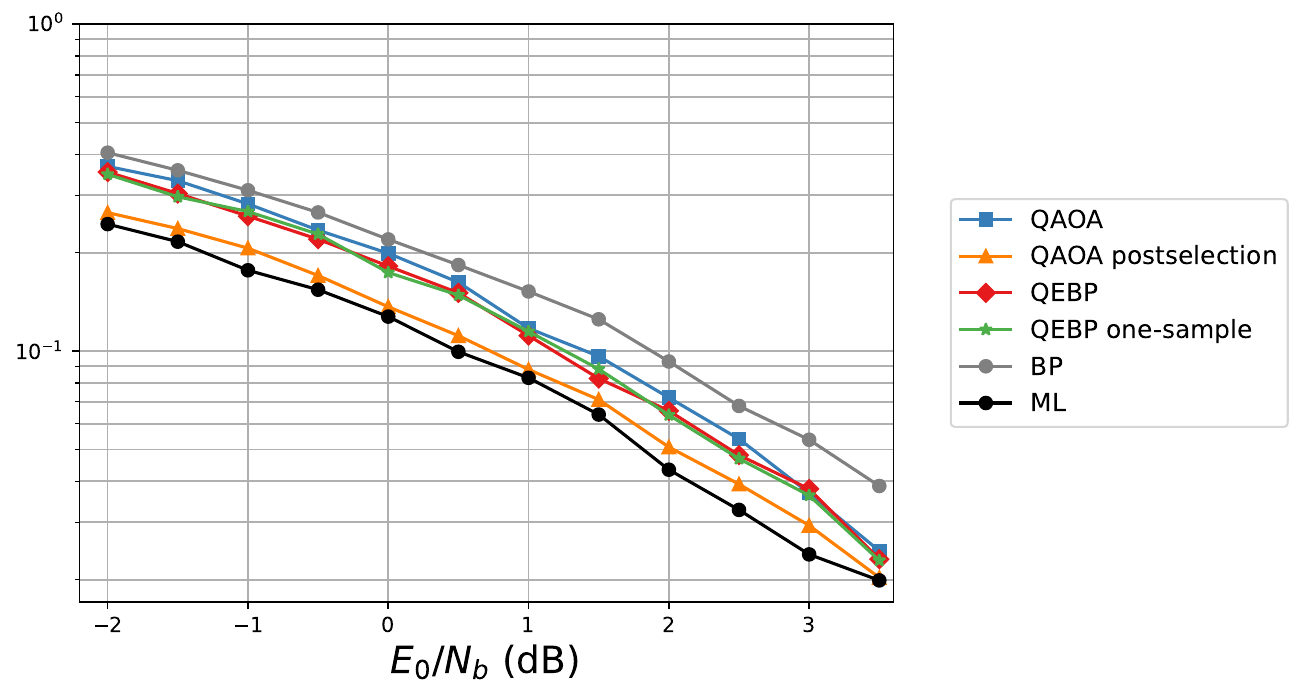}
  \caption{$[12,8]$-irregular code.}
  \label{fig:1285irr}
\end{subfigure}
\caption{Block error rate against signal-to-noise ratio $(E_0/N_b)$ for syndrome decoding with QAOA with $p=5$. We compare results using QAOA syndrome decoding, QAOA with codeword post-selection, and quantum-enhanced belief propagation decoding.}
\label{fig:1285}
\end{figure}

As discussed in \cref{section:zerotransmission}, without loss of generality, we assume the transmission of the all-zero codeword. For the QAOA syndrome decoder, we simulate the QAOA algorithm using the Julia library Yao \cite{Luo2020} and optimize the variational parameters using the BFGS algorithm with 10 random initialization points. For each algorithm run, we collect 10000 measurement samples and rank the output strings by number of appearances. The final solution is the highest-ranked string. The same ranking scheme is implemented for the post-selection algorithm on the strings that satisfy the codeword condition. In \cref{fig:1283}, we plot the BLER against the signal-to-noise ratio ($E_b/N_0$) for all versions of the decoding algorithm using QAOA with $p=3$. Similarly, in \cref{fig:1285}, we show our results for $p=5$. The results for $p \in \{1,\ldots,5\}$ are in \cref{section:all_layers}. 

For the irregular LDPC code (\cref{fig:1283irr} and \cref{fig:1285irr}) with $p=3$, the QAOA syndrome decoder with post-selection achieves a BLER of up to $72\%$ lower than QAOA syndrome decoding for $E_b/N_0 = \qty[mode = text]{3.5}{\decibel}$. However, for $p=5$ the largest improvement seen is of $30\%$ for $E_b/N_0 = \qty[mode = text]{0.5 }{\decibel}$. This shows that as QAOA produces better solutions, the impact of post-selecting in the BLER is reduced, as expected. A similar effect can be seen for the regular code and the other decoding protocols. 

The QEBP algorithm's BLER improvement on BP ranges over $13-40\%$ across both codes and levels. Remarkably, in \cref{fig:1285}, we can see how the QEBP algorithm initialized with one sample closely matches the QAOA syndrome decoding BLER for the regular code and the QEBP algorithm for the irregular one. In both instances, matching whichever approach has a lower BLER. This phenomenon may make it a good candidate over QEBP with the estimated probability of error as input.

\subsection{Average number of iterations before convergence}

One of the main characteristics affecting fast communications is the number of iterations it takes for belief propagation to converge to a solution \cite{Hailes2016}. We compare the number of iterations that it takes for the min-sum algorithm to converge to a solution, be it the correct one or not, and compare it with QEBP and QEBP with one sample as input.

\begin{figure}[ht]
\begin{subfigure}{.421\linewidth}
     \centering
    \includegraphics[width=\linewidth]{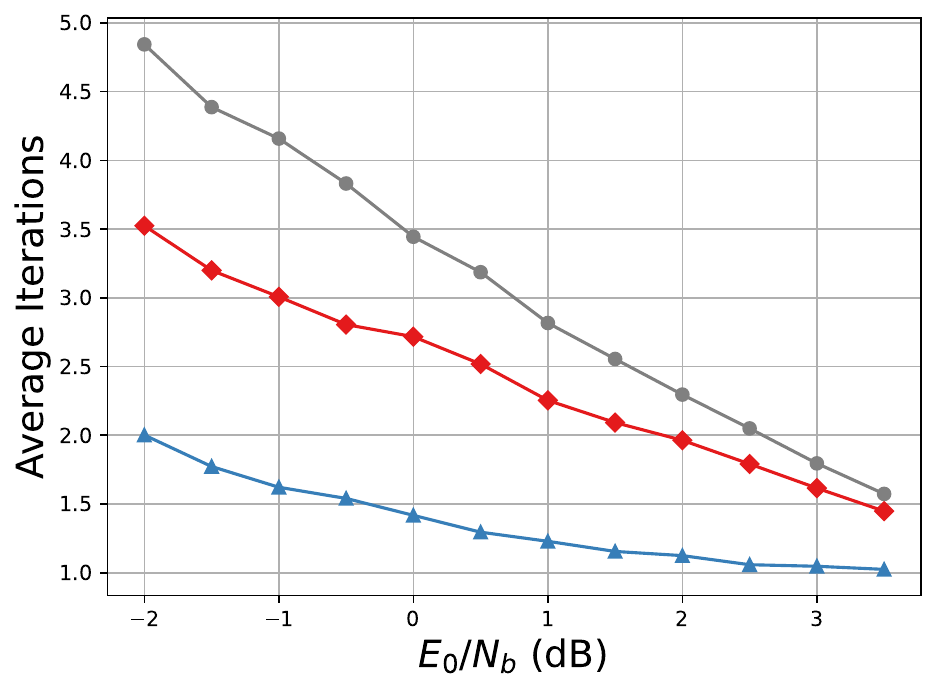}
    \caption{$[12,8]$-regular code.}
    \label{fig:iter1285}
\end{subfigure}
\begin{subfigure}{.579\linewidth}
\centering
    \includegraphics[width=\linewidth]{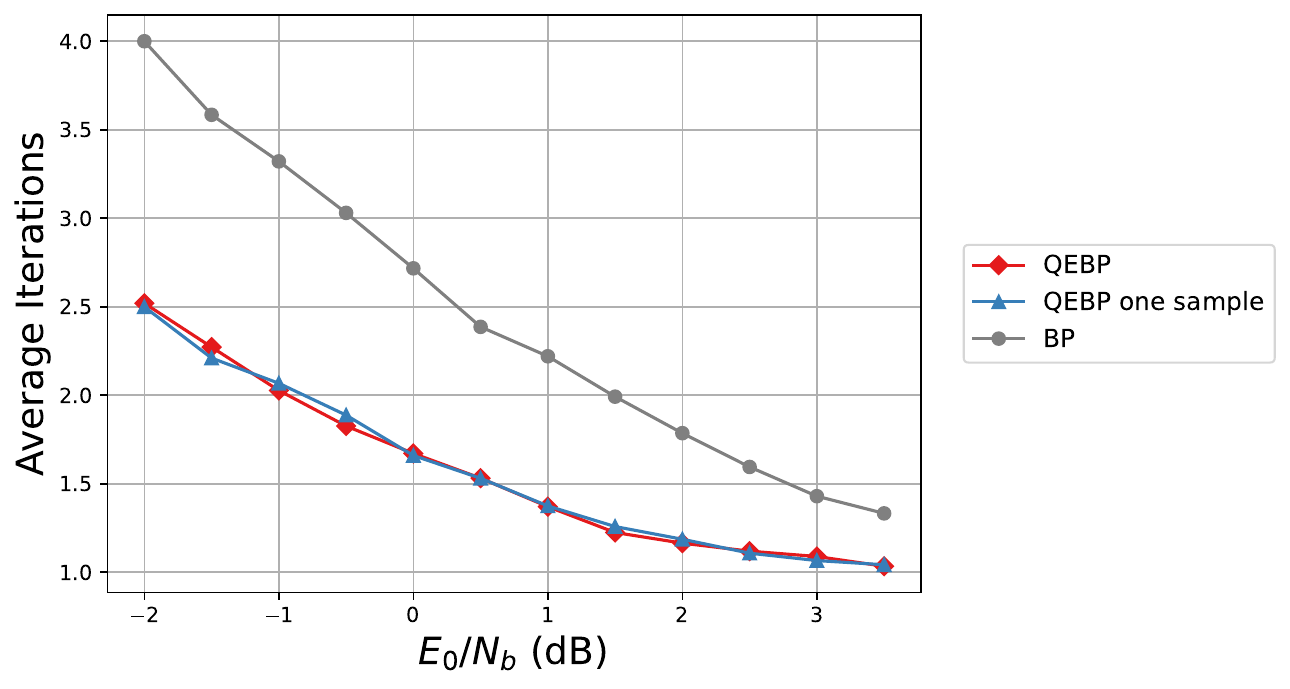}
    \caption{$[12,8]$-irregular code.}
    \label{fig:iter1285irr}
\end{subfigure}
\caption{Average number of iterations needed over 10000 decoding rounds for the belief propagation algorithm, QEBP and QEBP with a single sample as input.}
\label{fig:iters128}
\end{figure}

For the $(3,2)$-regular code, in \cref{fig:iter1285}, we find that the min-sum algorithm warm-started with the probability estimate from level 5 QAOA reduces the number of iterations needed for the convergence of belief propagation on average by $19.5\%$. The quantum-enhanced algorithm with the highest-count sample as input, however, is able to reduce the iterations by approximately $53.7\%$ on average.  Nonetheless, due to the quantum resources needed for each decoding round, the final processing throughput is smaller in the quantum-enhanced approach. In the irregular case, both methods display similar convergence, reducing iteration by $35\%$ on average.

%------------------------------------------------------------------
%------------------------------------------------------------------
%------------------------------------------------------------------

\section{QAOA syndrome decoding post-processing} \label{section:decode_protocol}
In this section, we analyze the decoding performance of 1-layer QAOA in the case of syndrome decoding for the $[n,1]$ repetition code, where one bit is repeated $n$ times. We consider three versions to obtain the error pattern from the QAOA circuit: one-sample QAOA, QAOA with post-selection, and QAOA with ranking. To compare these methods, we re-express the probability of correct decoding in terms of transfer matrices. Then, we use the eigenvalues of the transfer matrix to maximize the the probability of successful decoding.

\subsection{Probability of successful decoding} \label{section:prob_succ_rep}

Akin to the transfer matrix method used to calculate the free energy of the one-dimensional Ising model \cite{Kramers1941}, we want to express the probability of successful decoding for the repetition code in terms of the multiplication of repeated matrices. For the repetition code, the parity-check matrix has the following form
\begin{equation}
    H = \begin{pmatrix}
        1 & 1 & 0 & 0 &0 &0 & 0 \\
        0 & 1 & 1 & 0 &0 &0 & 0 \\
        0 & 0 & 1 & 1 &0 &0 & 0 \\
        0 & 0 & 0 & 1 &1 &0 & 0 \\
        0 & 0 & 0 & 0 &1 &1 & \ddots \\
    \end{pmatrix},
\end{equation}
with dimension $(n-1) \times n$. The corresponding syndrome-decoding QAOA cost Hamiltonian is
\begin{equation}
    \widehat{C}_{\mathrm{rep}}=\eta \sum_{j=1}^{n-1}\left(1-2 s_{j}\right) Z_{j} Z_{j+1} +\alpha \sum_{j=1}^{n} Z_{j}. \label{eq:hrep}
\end{equation}

Then, the probability of successful decoding, that is, the probability of obtaining the state $\Tilde{\boldsymbol{e}}$ with syndrome $\boldsymbol{s}$ when decoding a bit string $\boldsymbol{y}$ by measuring a 1-layer QAOA syndrome decoding circuit is given by

\begin{equation}
     \mathbb{P}_{\mathrm{QAOA}} (\Tilde{\boldsymbol{e}}|  \boldsymbol{s}= \boldsymbol{y} H^T, \alpha, \eta, \gamma, \beta) = 2^{-n} \left| \bra{\psi(-\beta)} \exp(-\mathrm{i}(-1)^{\Tilde{e}_n} \alpha \gamma Z)
    \overset{\curvearrowleft}{\prod_{k=1}^{n-1}} T_{\Tilde{e}_k}\ket{\psi(\beta)}  \right|^2, \label{eq:prob_decoding_ep}
\end{equation}
where the arrow on top of the product signifies multiplication on the left,
\begin{align*}
    T_{\Tilde{e}_k} =& \frac{1}{2}\left(\cos(\alpha \gamma) e^{-\mathrm{i}\beta} - \mathrm{i} (-1)^{\Tilde{e}_k}  \sin(\alpha \gamma) e^{\mathrm{i}\beta}\right)e^{-\mathrm{i} \gamma \eta} \mathbb{1} \\ & + \sqrt{-\frac{\mathrm{i}}{2}\sin(2\beta)} e^{\mathrm{i} \eta \gamma }\cos(\alpha \gamma) X \\ &
    - (-1)^{\Tilde{e}_k} \sqrt{-\frac{\mathrm{i}}{2}\sin(2 \beta)} e^{\mathrm{i} \eta \gamma }\sin(\alpha \gamma)Y \\ & + \frac{1}{2}\left(\cos(\alpha \gamma) e^{\mathrm{i}\beta} - \mathrm{i} (-1)^{\Tilde{e}_k}  \sin(\alpha \gamma) e^{-\mathrm{i}\beta}\right)e^{-\mathrm{i} \gamma \eta}Z,
\end{align*}
and $\ket{\psi(\beta)}= \begin{pmatrix}
        \sqrt{\cos\beta}\\
        \sqrt{-\mathrm{i}\sin\beta}
    \end{pmatrix}$. The full details of this derivation are given in \cref{section:prob_calcs}.

The matrix $T_{\Tilde{e}_k}$ has eigenvalues
\begin{equation}
\begin{split}
    \lambda_{\Tilde{e}_k,\pm} &= \frac{1}{2}\left(\cos(\alpha \gamma) e^{-\mathrm{i}\beta} - \mathrm{i} (-1)^{\Tilde{e}_k}  \sin(\alpha \gamma) e^{\mathrm{i}\beta}\right)e^{-\mathrm{i} \gamma \eta}\\ & \pm \frac{1}{2} \sqrt{-2\mathrm{i}\sin(2\beta) e^{\mathrm{i} \gamma  \eta} + \left(\cos^2\left(\alpha \gamma\right) e^{2\mathrm{i} \beta } - \sin^2\left(\alpha \gamma\right) e^{-2\mathrm{i} \beta } - \mathrm{i} (-1)^{\Tilde{e}_k}\sin(\alpha \gamma)\right)e^{-\mathrm{i} \gamma \eta}}, \label{eq:eigen_rep}
\end{split}
\end{equation}
which can be used to obtain the parameters $\beta$ and $\gamma$ that maximize the probability of successful decoding. 

The expected number of errors introduced by a BSC channel is given by
\begin{equation*}
    \mathbb{E}[|\boldsymbol{\Tilde{e}}|] = n \varepsilon,
\end{equation*}
where $\varepsilon$ is the crossover probability of the channel. In applications, the minimum SNR is around $\qty[mode = text]{20}{\decibel}$, resulting in a crossover probability very close to zero. Hence, for a string of length $n$, we should expect the number of errors to be low. Therefore, expecting the eigenvalues of $T_{\Tilde{e}_0}$ to dominate, we maximize $|\lambda_{0,+}|$  for $\beta$ and $\gamma$ setting the values for the balancing parameters to be $\alpha = 1$ and $\eta = 2$, as before. The result obtained from the optimization is $\gamma \approx 0.19419$ and $\beta \approx 0.506185$, which are the parameters that we will use in the following to compare post-processing protocols for the repetition code as the length of the codeword increases. These values for $\gamma$ and $\beta$ may be sub-optimal for some syndromes, as other terms in the expression for the probability of successful decoding in \cref{eq:prob_decoding_ep} may be, in turn, minimized. However, due to the similarity of the eigenvalues in \cref{eq:eigen_rep}, this effect is not likely to be, in general, prevalent.

\subsection{Block error rate for each post-processing strategy}

Now, we compare the expected block error rate for each post-processing strategy after transmission through the BSC channel. For a message of length $n$, the probability of receiving a message with $m$ errors after transmission through a BSC channel with crossover probability $\varepsilon$ is given by
\begin{equation}
    \mathbb{P}(m) = \varepsilon^m (1-\varepsilon)^{n-m}.
\end{equation}
We find the probability of wrong decoding for each decoding algorithm for information sent using BPSK modulation transmitted over an AWGN channel. In the following, we find the block error rate for each decoding protocol.

\subsubsection*{One-sample QAOA}
The expectation of wrong decoding by a single run of QAOA is given by
\begin{equation}
    \mathbb{E}[X]= \sum_{\Tilde{\boldsymbol{e}} \in \{0,1\}^n  \backslash \{\boldsymbol{0},\boldsymbol{1}\}}   \mathbb{P}(w_H(\Tilde{\boldsymbol{e}})) \left(1-\mathbb{P}_{\mathrm{QAOA}}\left(\Tilde{\boldsymbol{e}} |  \boldsymbol{s} = \boldsymbol{y} H^T\right)\right), \label{eq:osample}
\end{equation}
where $X$ represents the outcome of sending a message through the bit-flip channel; that is, $X$ gives the probability of obtaining a particular error string. Note that in \cref{eq:osample}, we have dropped the dependence on the balancing and variational parameters.

\subsubsection*{QAOA with post-selection}
The QAOA parity-check algorithm outputs an error string $\Tilde{\boldsymbol{e}}$, which added to the received message $\boldsymbol{y}$ should output a codeword $\boldsymbol{x}$. In this version of the decoding algorithm, all instances where $\boldsymbol{y} + \Tilde{\boldsymbol{e}}$ is not a codeword are thrown away. Therefore, for the $[n,1]$ repetition code, where there are only two possible codewords, the all-zero and the all-one strings, the probability of error is
\begin{equation}
    \mathbb{P}_{\text{error}}(\Tilde{\boldsymbol{e}}) = \frac{\mathbb{P}_{\text{QAOA}}\left(\Tilde{\boldsymbol{e}}'| \boldsymbol{s} = \boldsymbol{y} H^T\right)}{\mathbb{P}_{\text{QAOA}}\left(\Tilde{\boldsymbol{e}}| \boldsymbol{s} = \boldsymbol{y} H^T\right) + \mathbb{P}_{\text{QAOA}}\left(\Tilde{\boldsymbol{e}}'| \boldsymbol{s} = \boldsymbol{y} H^T\right)},
\end{equation}
where $\Tilde{\boldsymbol{e}}'$ is the complement error string to $\Tilde{\boldsymbol{e}}$, which would output the opposite result. The expectation of wrong decoding would be
\begin{equation}
    \mathbb{E}[X]= \sum_{\Tilde{\boldsymbol{e}} \in \{0,1\}^n \backslash \{\boldsymbol{0},\boldsymbol{1}\}} \mathbb{P}(w_H(\Tilde{\boldsymbol{e}}))\mathbb{P}_{\text{error}} (\Tilde{\boldsymbol{e}}).
\end{equation}

\subsubsection*{QAOA with ranking}
 In this approach, the QAOA parity-check algorithm is run multiple times to build an estimate of the probability of obtaining each string. The solutions obtained are then ranked with respect to their corresponding output probability. Define $r_{\Tilde{\boldsymbol{e}}}$ as follows
\begin{equation}
    r_{\Tilde{\boldsymbol{e}}} = \begin{cases}
        r_{\Tilde{\boldsymbol{e}}} = 1 \hspace{1cm} \Tilde{\boldsymbol{e}} \neq \Tilde{\boldsymbol{e}}',\\
        r_{\Tilde{\boldsymbol{e}}} = 0 \hspace{1cm} \Tilde{\boldsymbol{e}} = \Tilde{\boldsymbol{e}}',
    \end{cases}
\end{equation}
that is, $r_{\Tilde{\boldsymbol{e}}}$ is one when the first string in the ranking differs from the expected error string for the input syndrome $\Tilde{\boldsymbol{e}}'$ and zero when the correct string is found. 

The expectation of wrong syndrome decoding, in this case, is given by
\begin{equation}
    \mathbb{E}[X]= \sum_{\Tilde{\boldsymbol{e}} \in \{0,1\}^n \backslash \{\boldsymbol{0},\boldsymbol{1}\}} \mathbb{P}(w_H(\Tilde{\boldsymbol{e}}))   r_{\Tilde{\boldsymbol{e}}}.
    %\mathbb{P}(\Tilde{\boldsymbol{e}}'| \Tilde{y}), %d_H(\Tilde{\boldsymbol{e}},\Tilde{\boldsymbol{e}}'),
\end{equation}
To estimate $r_{\Tilde{\boldsymbol{e}}}$, we find the expected number of appearances for each string after $r$ rounds. Let $\sigma_{\Tilde{\boldsymbol{e}}}$ be the number of times string $\Tilde{\boldsymbol{e}}$ is obtained after $r$ rounds. Then, the expectation of $\sigma_{\Tilde{\boldsymbol{e}}}$ is
\begin{equation}
    \mathbb{E}[\sigma_{\Tilde{\boldsymbol{e}}}] = r \mathbb{P}_{\mathrm{QAOA}}\left(\Tilde{\boldsymbol{e}} |  \boldsymbol{s} = \boldsymbol{y} H^T\right),
\end{equation}
where $\boldsymbol{s}$ is the syndrome obtained from the correct error string $\Tilde{\boldsymbol{e}}'$.

We compare these post-processing strategies with majority vote decoding, the optimal decoding strategy for the repetition code. For majority vote decoding, the block error rate of the repetition code with $n$ repetitions over a bit-flip channel with cross-over probability $\varepsilon$ is given by
\begin{equation}
    \mathrm{BLER} = \sum_{m = (n+1)/2}^n {\binom{n}{m}} \varepsilon^m (1-\varepsilon)^{n-m}.
\end{equation}

\begin{figure}
    \centering 
    \includegraphics[width=0.8\textwidth]{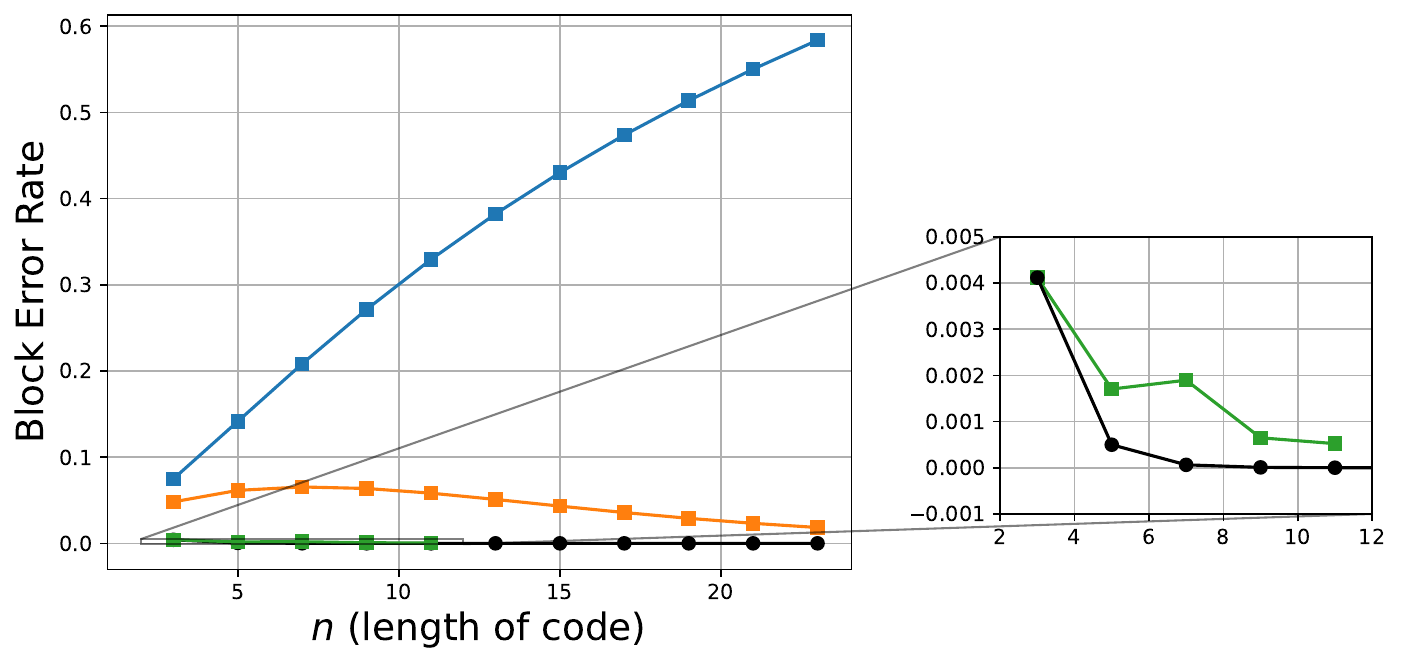}   
    \caption{Block error rate against code length for $E_b/N_0 = \qty[mode = text]{2}{\decibel}$. We compare majority vote (\protect\tikz[baseline=-0.5ex]{\protect\draw[thick,black] (-.3,0) -- (.3,0); \protect\draw[black,fill] (0,0) circle (2pt);}) with the different QAOA post-processing approaches: one-sample QAOA (\protect\tikz[baseline=-0.5ex]{\protect\draw[thick,matblue] (-.3,0) -- (.3,0); \protect\draw[matblue,fill] (-2pt,-2pt) rectangle (2pt,2pt);}), QAOA with ranking of 1000 samples (\protect\tikz[baseline=-0.5ex]{\protect\draw[thick,matgreen] (-.3,0) -- (.3,0); \protect\draw[matgreen,fill] (-2pt,-2pt) rectangle (2pt,2pt);}) and QAOA with post-selection (\protect\tikz[baseline=-0.5ex]{\protect\draw[thick,matorange] (-.3,0) -- (.3,0); \protect\draw[matorange,fill] (-2pt,-2pt) rectangle (2pt,2pt);}).}
    \label{fig:block_snr_rep_code}
\end{figure}

\begin{figure}
    \centering
    \includegraphics[width=0.5\linewidth]{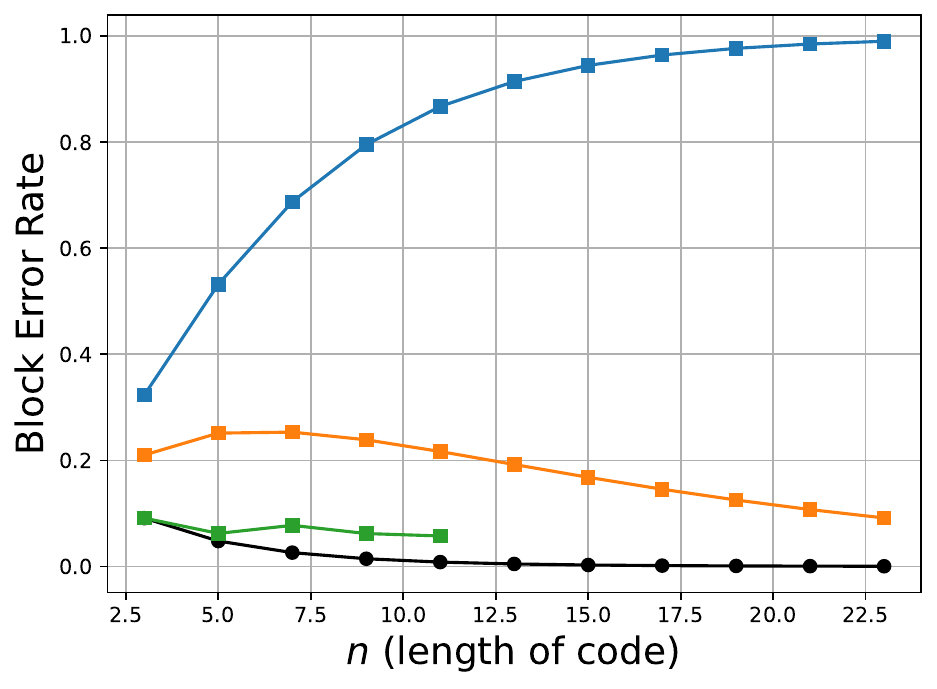}
    \caption{Block error rate against code length for $E_b/N_0 = \qty[mode = text]{-4}{\decibel}$. We compare majority vote (\protect\tikz[baseline=-0.5ex]{\protect\draw[thick,black] (-.3,0) -- (.3,0); \protect\draw[black,fill] (0,0) circle (2pt);}) with the different QAOA post-processing approaches: one-sample QAOA (\protect\tikz[baseline=-0.5ex]{\protect\draw[thick,matblue] (-.3,0) -- (.3,0); \protect\draw[matblue,fill] (-2pt,-2pt) rectangle (2pt,2pt);}), QAOA with ranking of 1000 samples (\protect\tikz[baseline=-0.5ex]{\protect\draw[thick,matgreen] (-.3,0) -- (.3,0); \protect\draw[matgreen,fill] (-2pt,-2pt) rectangle (2pt,2pt);}) and QAOA with post-selection (\protect\tikz[baseline=-0.5ex]{\protect\draw[thick,matorange] (-.3,0) -- (.3,0); \protect\draw[matorange,fill] (-2pt,-2pt) rectangle (2pt,2pt);}).}
    \label{fig:block_snr_rep_code-4}
\end{figure}

\Cref{fig:block_snr_rep_code,fig:block_snr_rep_code-4} show the block error rate for odd values of $n$ in the interval from 3 to 23; odd values were chosen to ensure a definite response to the majority vote decoding. In \cref{fig:block_snr_rep_code-4}, for $E_b/N_0 = \qty[mode = text]{-4}{\decibel}$, rapidly approaches one, providing worse than random guessing error performance for $n>3$. We note that discarding the rounds that produce strings that are not part of the code drastically improves the BLER scaling, converging to a BLER close to zero for large $n$ when $E_b/N_0 = \qty[mode = text]{0}{\decibel}$. However, as the code size increases, the probability of outputting a codeword decreases, which may make the method intractable for large $n$ due to its high computational cost. By estimating the probability distribution through sampling, we can reduce the BLER significantly. The results shown are for an estimate of 1000 samples. However, the resulting BLER does not converge to zero as with maximum-likelihood decoding.

By optimizing $\gamma$ and $\beta$ using the transfer method described in \cref{section:prob_succ_rep}, we could determine unique parameters for level-1 QAOA that achieve close to maximum-likelihood decoding performance for positive $E_b/N_0$. In the case of one-sample QAOA, we obtain a linear scaling for positive $E_b/N_0$, which is improved by post-processing, indicating that, in general, post-processing may be required to get good solutions from QAOA.

\subsection{Expected number of QAOA decoding rounds}

\begin{figure}
    \centering
    \includegraphics[width=.5\textwidth]{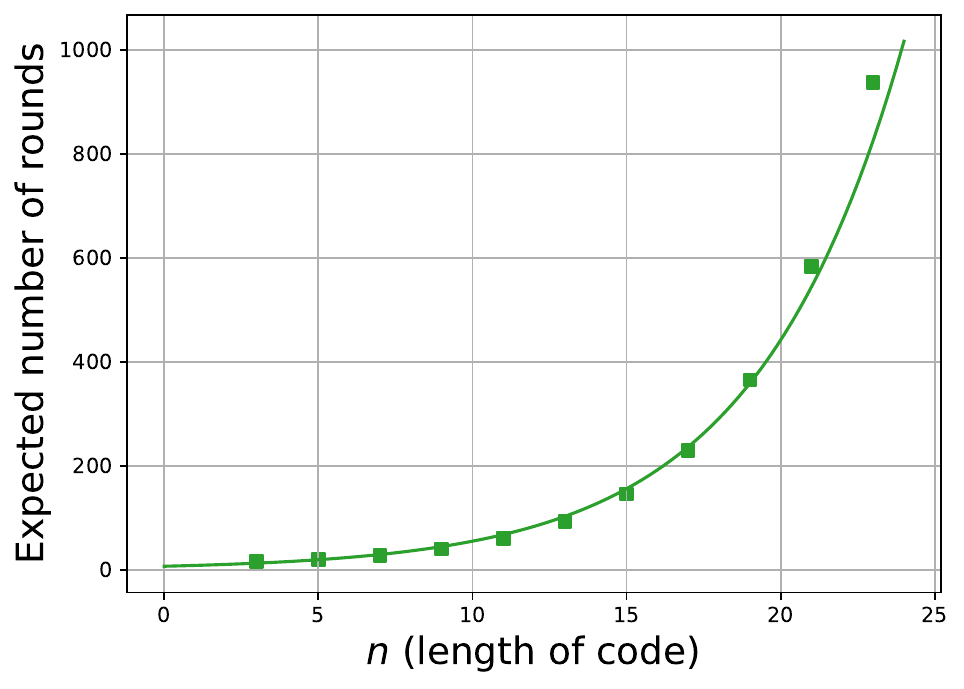}
    
    \caption{Rounds required for the post-selected QAOA decoding algorithm to terminate against the length of the repetition code for $E_b/N_0 = \qty[mode = text]{2}{\decibel}$.}
    \label{fig:term_snr_rep_code}
\end{figure}

\begin{figure}[ht]
    \centering
      \includegraphics[width=.7\textwidth]{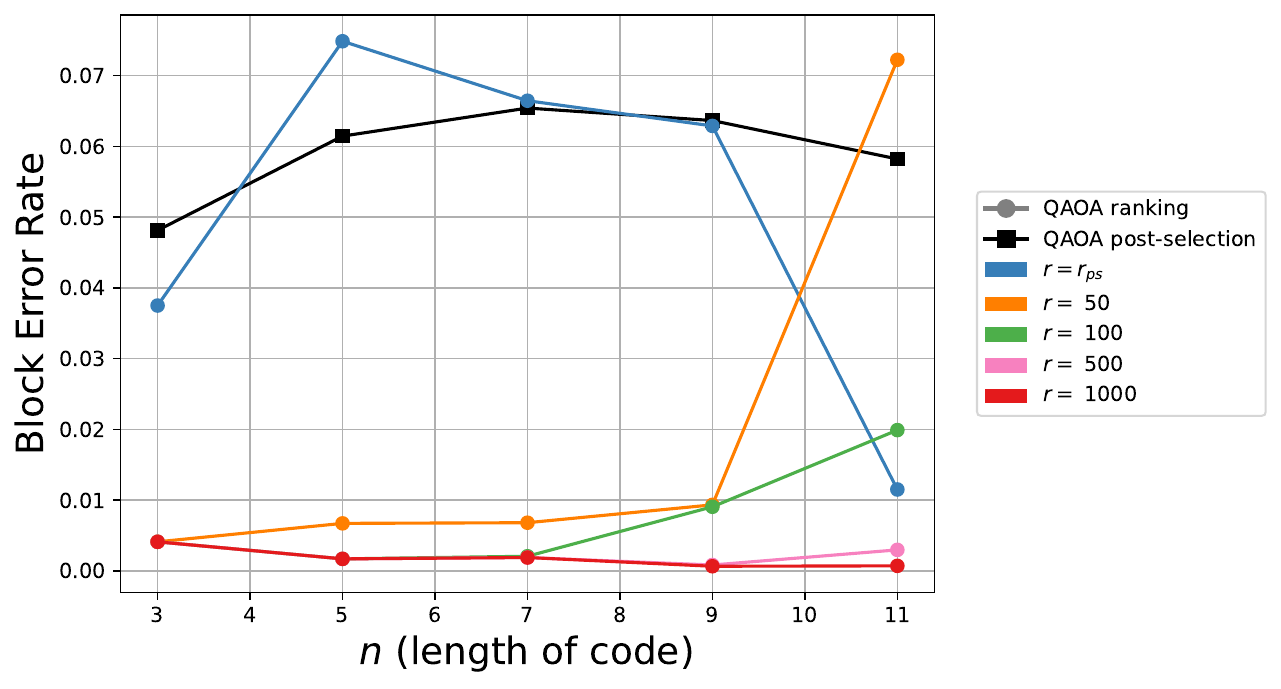}
    \caption{Block error rate against length of code $n$ for different number of rounds $r$ for the QAOA with ranking protocol. We show results for $E_b/N_0 = \qty[mode = text]{2}{\decibel}$ and compare them to the post-selected QAOA algorithm.}
    \label{fig:mr}
\end{figure}

To reach a solution in the case of QAOA decoding with the post-selection algorithm, the corrected string needs to be a codeword. The probability of outputting a codeword after a single round of QAOA decoding is given by
\begin{equation}
    \mathbb{P}_{\mathrm{codeword}} \left(\Tilde{\boldsymbol{e}}) = \mathbb{P}_\mathrm{QAOA}(\Tilde{\boldsymbol{e}} |  \boldsymbol{s} = \boldsymbol{y} H^T\right) + \mathbb{P}_\mathrm{QAOA}\left(\Tilde{\boldsymbol{e}}' |  \boldsymbol{s} = \boldsymbol{y} H^T\right),
\end{equation}
where $\Tilde{\boldsymbol{e}}'$ is the complement error string to $\Tilde{\boldsymbol{e}}$, as before. Then, the expected termination probability would be
\begin{equation}
    \mathbb{P}_{\mathrm{termination}} = \mathbb{E}[\mathbb{P}_{\text{codeword}} (\Tilde{\boldsymbol{e}})] =\sum_{\Tilde{\boldsymbol{e}} \in \{0,1\}^n} \mathbb{P}(w_H(\Tilde{\boldsymbol{e}}))\mathbb{P}_{\text{codeword}} (\Tilde{\boldsymbol{e}}),
\end{equation}
and the expected number of rounds needed for the QAOA decoding algorithm with post-selection to terminate is given by
\begin{equation}
    \mathbb{E}[\mathrm{rounds}] = \frac{1}{\mathbb{P}_{\mathrm{termination}}}.
\end{equation}
In \cref{fig:term_snr_rep_code}, we observe that the expected number of rounds for the algorithm to terminate increases exponentially with the length of the code, as expected. The fit equation is $2^{ax+b}$, where $a=0.3005264440307172$ and $b=2.7767955365873327$. In \cref{fig:mr}, we compare the block error rate for the QAOA ranking protocol for a varying number of total rounds and contrast it with the post-selected QAOA algorithm. In the post-selected case, the expected number of rounds needed for a code length in the range in the figure, that is, from three to eleven, ranges from 8 to 36. In blue, we compare the results obtained by setting the number of rounds in the QAOA with ranking case to be the expected number of rounds needed for the post-selected case to terminate. As shown in the figure, the BLER varies for each code length, making it difficult to draw conclusions from this comparison. In practice, the low number of rounds makes the final ranking of solutions strongly dependent on the particular experiment, therefore requiring a higher number of samples for stable results. This effect can be appreciated when looking at the results for an increasing number of rounds. For smaller code lengths, the probability of successful decoding is higher, thus requiring fewer rounds for the BLER to stabilize. Overall, the QAOA ranking method requires more rounds on average to produce stable results than the post-selection counterpart. However, it can achieve BLERs more than five times lower. This reduces the question of which methods to use to the question of how many rounds one is able to perform.

\section{Conclusion} \label{section:conclusion}
We have introduced the quantum-enhanced belief propagation algorithm, which uses QAOA to warm-start belief propagation. By performing exact simulations for regular and irregular LDPC codes, we have shown that QEBP achieves lower BLER than belief propagation for small block sizes and helps it converge to a solution in fewer iterations. In practice, however, the code block sizes used in error correction with LDPC codes are in the order of thousands, with the maximum code block size set by 3GPP being 8448. The current generation of Noisy Intermediate-Scale Quantum (NISQ) devices prevents us from implementing such large code sizes reliably, as they are susceptible to noise, which compromises the quality of the solution.

While this paper does not explore the hardware implementation of the QEBP decoder, it is a critical consideration for its realization. Unlike the belief propagation algorithm, which, when implemented using field-programmable gate array (FPGA) devices, can accommodate different families of codes without reprogramming \cite{Choi2016}, QAOA requires re-optimizing the parameters of the quantum circuit for each decoding round. This leads to a lack of decoding time flexibility, which could potentially slow down the communication process. Nonetheless, the QEBP algorithm's potential advantage lies in its modular integration into existing decoding architecture. It allows the quantum device to be re-programmed while the classical decoding architecture continues working and is only added to the routine once it is ready. As we have done for the repetition code, there is also the possibility of speeding up the parameter optimization routine by finding shared parameters between syndromes in advance. 

We have also investigated how different QAOA post-processing strategies may affect the BLER as the code size increases by looking at the repetition code. In particular, as the code length increases, the probability of successful decoding decreases exponentially with the code length. This makes the BLER for the single-round QAOA decoding approach rapidly reach one as the code length increases. Through post-selection and repeated sampling, we reduced the scaling of the BLER. Yet, in contrast with maximum-likelihood decoding, neither method is shown to converge to zero with the code length. Then, the question of which method is preferred is reduced to how many decoding rounds one can perform. In the case where performing a decoding round has a high cost associated with it, then the codeword post-processing strategy produces more stable results, although the number of rounds increases exponentially with the code length. However, if one is able to conduct more rounds, the ranking method provides better error-correcting performance. 
% \am{Do we want to say something stronger here about the level of performance that we found for the repetition code? In particular, that the success probability was exponentially small? I don't think this is discussed above -- maybe we can discuss when we meet.}

Further work is needed to understand how the algorithm can perform in practice. One potential avenue of study is to carry out noisy simulations to gauge how resilient the QEBP algorithm is against noise in the QAOA circuit. Another possible direction is to investigate convergence in belief propagation and how the initial guess affects the trapping sets of the code.

\section*{Acknowledgments}
S.M.P.G. is supported by the Engineering and Physical Sciences Research Council (EPSRC) [grant no.\ EP/S021582/1]. AM  acknowledges funding from the European Research Council (ERC) under the European Union’s Horizon 2020 research and innovation programme (grant agreement No.\ 817581).

% \newpage
\section*{References}
\printbibliography[heading = none]
\newpage
\begin{appendices}
    \appendix
    
\section{Calculations for probability of successful decoding by QAOA syndrome decoding algorithm} \label{section:prob_calcs}

In this section, we show how to re-write the probability of successful decoding using transfer matrices, as given in \cref{eq:prob_decoding_ep}. Let $\widehat{C} =  \eta \sum_{j=1}^{n-1}(1-2s_j) Z_j Z_{j + 1} + \alpha \sum_{j=1}^{n-1} Z_{j}$. We first re-express the amplitude of a state $\boldsymbol{\Tilde{e}}$ state with syndrome $\boldsymbol{s}$ as
\begin{align*}
    &\bra{\boldsymbol{\Tilde{e}}} \exp(-\mathrm{i}\beta \sum_{j=1}^n X_j)\exp(-\mathrm{i}\gamma \left( \eta \sum_{j=1}^{n-1} (1-2s_j) Z_j Z_{j + 1} + \alpha \sum_{j=1}^{n} Z_{j} \right))\ket{+}^{\otimes n}\\
    & = \bra{\boldsymbol{\Tilde{e}}}\exp(-\mathrm{i}\beta \sum_{j=1}^n X_j)\exp(-\mathrm{i}\gamma \widehat{C}) \exp(-\mathrm{i}\gamma \alpha Z_n)\ket{+}^{\otimes n}\\
    & = \bra{\boldsymbol{\Tilde{e}}}\exp(-\mathrm{i}\beta \sum_{j=1}^n X_j)\exp(-\mathrm{i}\gamma \widehat{C}) \cos(\alpha \gamma)\ket{+}^{\otimes n}\\ & - \bra{\Tilde{e}}\exp(-\mathrm{i}\beta \sum_{j=1}^n X_j)\exp(-\mathrm{i}\gamma \widehat{C}) \mathrm{i} \sin(\alpha \gamma)Z_n\ket{+}^{\otimes n}.
\end{align*}

The first term in the previous expression can be re-written by inserting a closure relation as

\begin{align*}
    &\cos(\alpha \gamma) \bra{\Tilde{e}}\exp(-\mathrm{i}\beta\sum_{j=1}^n X_j)\exp(-\mathrm{i} \gamma \widehat{C})\ket{+}^{\otimes n}\\
    & = \cos(\alpha \gamma)\bra{\Tilde{e}}\exp(-\mathrm{i}\beta \sum_{j=1}^n X_j)\left(\sum_{z \in \{0, 1\}^n}\ket{z}\bra{z}\right)  \exp(-\mathrm{i} \gamma \widehat{C})\ket{+}^{\otimes n}\\
    & = \cos(\alpha \gamma)2^{-n/2}\sum_{z \in \{0, 1\}^n}\left(\cos \beta \right)^{n - |\Tilde{e}+ z|}\left(-\mathrm{i}\sin\beta\right)^{|\Tilde{e} + z|}\exp(-\mathrm{i}\gamma C), \\
    %
    %
    %
    % & = \cos(\alpha \gamma)2^{-n/2}\sum_{z \in \{0, 1\}^n}\left(\cos \beta \right)^{n - |\Tilde{e}+ z|}\left(-\mathrm{i}\sin\beta\right)^{|\Tilde{e} + z|}e^{-\mathrm{i}\gamma \left(\eta \sum_{j=1}^{n-1} (-1)^{z_j + z_{j + 1}} (-1)^{\Tilde{e}_iH_{ij}}+\alpha \sum_{j=1}^{n-1} (-1)^{z_j}\right)} \\
    %
    %
    %
    % & = \cos(\alpha \gamma)2^{-n/2}\sum_{z \in \{0, 1\}^n}\left(\cos \beta \right)^{n - |\Tilde{e}+ z|}\left(-\mathrm{i}\sin\beta\right)^{|\Tilde{e} + z|}e^{-\mathrm{i}\gamma \left(\eta \sum_{j=1}^{n-1} (-1)^{z_j + z_{j + 1}} (-1)^{\Tilde{e}_j + \Tilde{e}_{j+1}}+\alpha \sum_{j=1}^{n-1} (-1)^{z_j}\right)},
\end{align*}
where 
\begin{equation*}
    C = \eta \sum_{j=1}^{n-1} (-1)^{z_j + z_{j + 1} + s_j}+\alpha \sum_{j=1}^{n-1} (-1)^{z_j} = \eta \sum_{j=1}^{n-1} (-1)^{z_j + z_{j + 1}} (-1)^{\Tilde{e}_j + \Tilde{e}_{j+1}}+\alpha \sum_{j=1}^{n-1} (-1)^{z_j}
\end{equation*}
and we have used the fact that $s_j = \Tilde{e}_{i}H_{ij} = \Tilde{e}_{j}+\Tilde{e}_{j+1}$. Now, we perform the following change of variables $z' = z + \Tilde{e}$ such that

\begin{equation*}
    C = \eta \sum_{j=1}^{n-1} (-1)^{z'_j + z'_{j + 1}} +\alpha \sum_{j=1}^{n-1} (-1)^{z'_j + \Tilde{e}_j},
\end{equation*}
and
\begin{align*}
    & \cos(\alpha \gamma)2^{-n/2}\sum_{z' \in \{0, 1\}^n}\left(\cos \beta \right)^{n - |z'|}\left(-\mathrm{i}\sin\beta\right)^{|z'|}\exp(-\mathrm{i}\gamma C)\\
    & = \cos(\alpha \gamma)2^{-n/2} \begin{pmatrix}
        \sqrt{\cos\beta}\\
        \sqrt{-\mathrm{i}\sin\beta}
    \end{pmatrix}^T  \overset{\curvearrowleft}{\prod_{k=1}^{n-1}} \left\{\begin{pmatrix}
        \sqrt{\cos\beta} & 0\\
        0 & \sqrt{-\mathrm{i}\sin\beta}
    \end{pmatrix} \begin{pmatrix}
        e^{-\mathrm{i} \eta \gamma } & e^{ \mathrm{i} \eta \gamma }\\
        e^{ \mathrm{i} \eta \gamma } & e^{-\mathrm{i} \eta \gamma }
    \end{pmatrix} \right.\\
    &\left. \ \begin{pmatrix}
        e^{\mathrm{i}(-1)^{\Tilde{e}_k +1} \alpha \gamma } & 0\\
        0 & e^{ \mathrm{i}(-1)^{\Tilde{e}_k}  \alpha \gamma }
    \end{pmatrix}
    \begin{pmatrix}
        \sqrt{\cos\beta} & 0\\
        0 & \sqrt{-\mathrm{i}\sin\beta}
    \end{pmatrix}\right\}
    \begin{pmatrix}
        \sqrt{\cos\beta}\\
        \sqrt{-\mathrm{i}\sin\beta}
    \end{pmatrix} \\
    & = \cos(\alpha \gamma)2^{-n/2}\begin{pmatrix}
        \sqrt{\cos\beta}\\
        \sqrt{-\mathrm{i}\sin\beta}
    \end{pmatrix}^T \\
    &  \overset{\curvearrowleft}{\prod_{k=1}^{n-1}} \left\{ \begin{pmatrix}
          \cos \beta e^{-\mathrm{i}\gamma (\eta + (-1)^{\Tilde{e}_k} \alpha)}& \sqrt{- \frac{\mathrm{i}}{2} \sin (2\beta)} e^{-\mathrm{i}\gamma (-\eta + (-1)^{\Tilde{e}_k+1} \alpha)}\\
          \sqrt{- \frac{\mathrm{i}}{2} \sin ( 2\beta)} e^{-\mathrm{i}\gamma (-\eta + (-1)^{\Tilde{e}_k} \alpha)} & -\mathrm{i} \sin \beta e^{-\mathrm{i}\gamma (\eta + (-1)^{\Tilde{e}_k +1} \alpha)}
    \end{pmatrix}\right\}\begin{pmatrix}
        \sqrt{\cos\beta}\\
        \sqrt{-\mathrm{i}\sin\beta}
    \end{pmatrix}.
\end{align*}
Here, we label the matrix inside the curly brackets as $T_{\Tilde{e}_k}$.
% Choose the matrix inside the brackets to be $T_{\Tilde{e}_k}$. In the case where the syndrome is the zero vector, we had that 
% \begin{equation}
%     \overset{\curvearrowleft}{\prod_{k=1}^{n-1}} T_{\Tilde{e}_k} = T_0^{n-1}.
% \end{equation}

Obtaining a similar expression for the other term and using $\ket{\psi(\beta)}= \begin{pmatrix}
        \sqrt{\cos\beta}\\
        \sqrt{-\mathrm{i}\sin\beta}
    \end{pmatrix}$, the amplitude becomes

\begin{align*}
    &\bra{\Tilde{e}}\exp(-\mathrm{i}\beta \sum_{j=1}^n X_j)\exp(-\mathrm{i}\gamma \left( \eta \sum_{j=1}^{n-1} (1-2s_j) Z_j Z_{j + 1} + \alpha \sum_{j=1}^{n} Z_{j} \right))\ket{+}\\
    & = 2^{-n/2} ( \cos(\alpha \gamma)\bra{\psi(-\beta)} 
    \overset{\curvearrowleft}{\prod_{k=1}^{n-1}} T_{\Tilde{e}_k}\ket{\psi(\beta)}
     -\mathrm{i}(-1)^{\Tilde{e}_n}\sin(\alpha \gamma)\bra{\psi(-\beta)} Z 
    \overset{\curvearrowleft}{\prod_{k=1}^{n-1}} T_{\Tilde{e}_k}\ket{\psi(\beta)} )\\
    & = 2^{-n/2} \bra{\psi(-\beta)} \left(\cos(\alpha \gamma) 
      -\mathrm{i}(-1)^{\Tilde{e}_n}\sin(\alpha \gamma) Z \right) 
    \overset{\curvearrowleft}{\prod_{k=1}^{n-1}} T_{\Tilde{e}_k}\ket{\psi(\beta)} \\
    & = 2^{-n/2} \bra{\psi(-\beta)} \exp(-\mathrm{i}(-1)^{\Tilde{e}_n} \alpha \gamma Z)
    \overset{\curvearrowleft}{\prod_{k=1}^{n-1}} T_{\Tilde{e}_k}\ket{\psi(\beta)}.
\end{align*}

Therefore, we obtain that the probability of successful decoding of the repetition code by the QAOA syndrome decoding algorithm is

\begin{equation}
     \mathrm{P}_{\rm{QAOA}} (\Tilde{\boldsymbol{e}}|  \boldsymbol{s}= \boldsymbol{y} H^T, \alpha, \eta, \gamma, \beta) = 2^{-n} \left| \bra{\psi(-\beta)} \exp(-\mathrm{i}(-1)^{\Tilde{e}_n} \alpha \gamma Z)
    \overset{\curvearrowleft}{\prod_{k=1}^{n-1}} T_{\Tilde{e}_k}\ket{\psi(\beta)}  \right|^2.
\end{equation}

\section{LDPC codes used for simulations.}
The parity-check matrix corresponding to the $[12,8]$-irregular LPDC code used in \cref{section:bpqaoa} is
\begin{equation}
     \begin{pmatrix}
        1 & 0 & 0 & 0 & 0 & 0 & 0 & 0 & 0 & 0 & 1 & 0  \\
        0 & 0 & 0 & 1 & 0 & 0 & 1 & 0 & 0 & 0 & 0 & 1  \\
        0 & 1 & 1 & 0 & 0 & 1 & 0 & 0 & 0 & 0 & 0 & 0  \\
        1 & 1 & 0 & 0 & 0 & 0 & 1 & 0 & 0 & 0 & 0 & 0  \\
        0 & 0 & 0 & 0 & 1 & 0 & 0 & 1 & 1 & 0 & 0 & 0  \\
        0 & 0 & 0 & 1 & 0 & 0 & 0 & 1 & 0 & 1 & 0 & 0  \\
        0 & 0 & 0 & 0 & 0 & 0 & 0 & 0 & 1 & 1 & 0 & 1  \\
        0 & 0 & 1 & 0 & 1 & 1 & 0 & 0 & 0 & 0 & 0 & 0  
\end{pmatrix}.  \label{eq:128irregular}
\end{equation}
The $(2,3)$-regular LDPC code is
\begin{equation}
    \begin{pmatrix}
        0 & 0 & 0 & 1 & 0 & 0 & 1 & 0 & 1 & 0 & 0 & 0  \\
        1 & 0 & 0 & 0 & 0 & 0 & 0 & 1 & 0 & 0 & 1 & 0  \\
        0 & 0 & 0 & 1 & 0 & 1 & 0 & 0 & 0 & 0 & 0 & 1  \\
        0 & 1 & 0 & 0 & 1 & 0 & 0 & 0 & 0 & 1 & 0 & 0  \\
        1 & 0 & 0 & 0 & 1 & 0 & 0 & 1 & 0 & 0 & 0 & 0  \\
        0 & 1 & 1 & 0 & 0 & 0 & 0 & 0 & 1 & 0 & 0 & 0  \\
        0 & 0 & 1 & 0 & 0 & 0 & 1 & 0 & 0 & 0 & 1 & 0  \\
        0 & 0 & 0 & 0 & 0 & 1 & 0 & 0 & 0 & 1 & 0 & 1 
            \end{pmatrix}. \label{eq:128regular}
\end{equation}

    \section{Error performance for QAOA layers 1-5} \label{section:all_layers}
\begin{figure}[ht]
\begin{subfigure}{.5\linewidth}
    \centering
      \includegraphics[width=\linewidth]{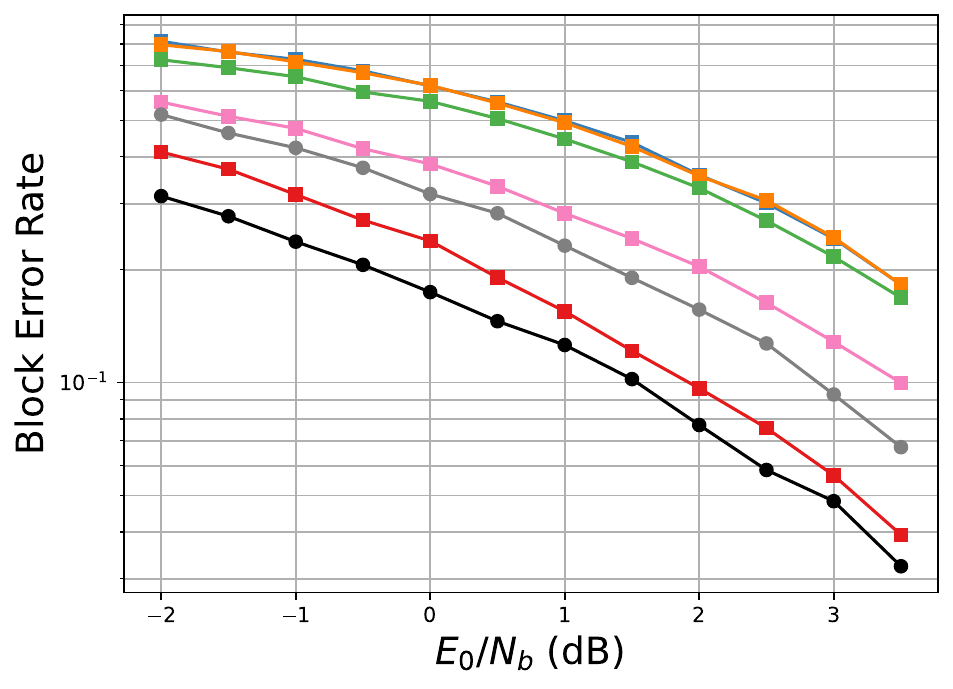}
      \caption{Syndrome decoding with QAOA.}
      \label{fig:qaoa128ap}
\end{subfigure}%
\begin{subfigure}{.5\linewidth}
\centering
  \includegraphics[width=\linewidth]{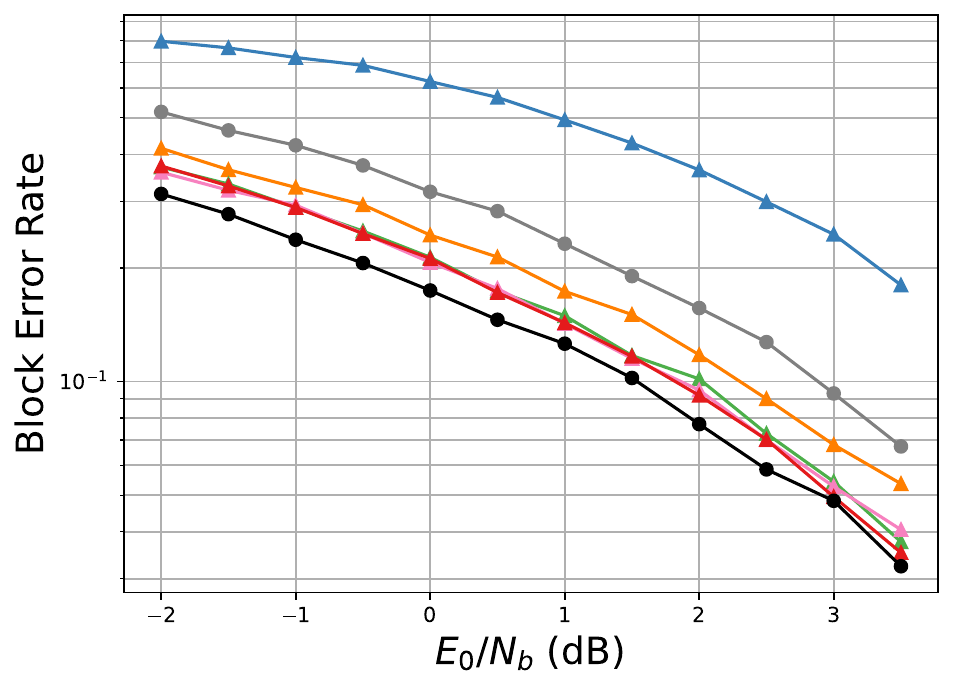}
  \caption{Decoding with codeword post-selection.}
  \label{fig:qaoapost128ap}
\end{subfigure}\\[1ex]
\begin{subfigure}{.5\linewidth}
    \centering
    \includegraphics[width=\linewidth]{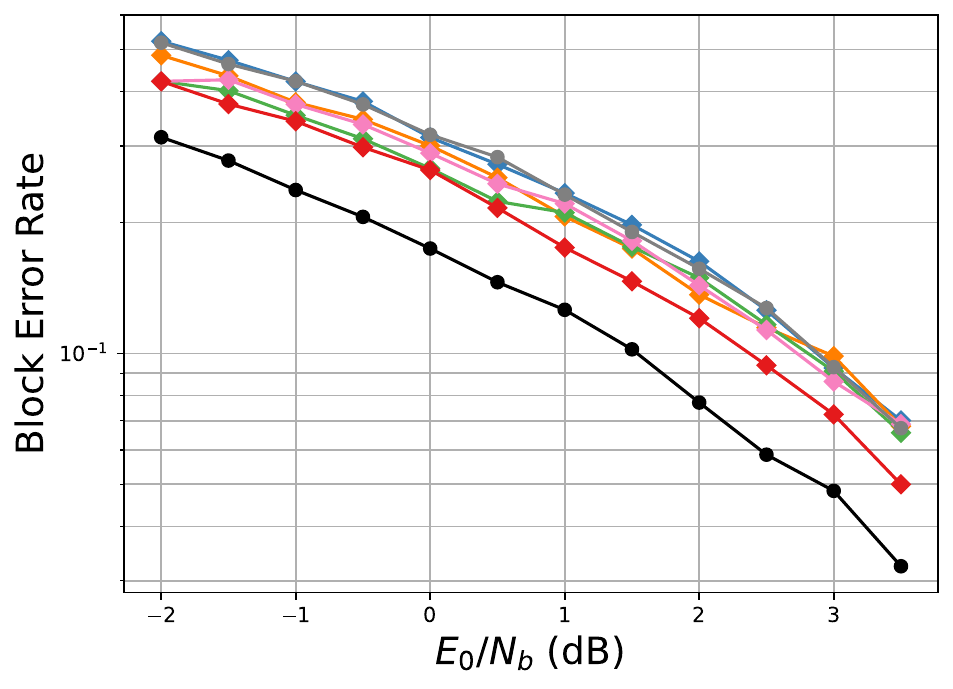}
  \caption{Decoding with QAOA-enhanced belief propagation.}
  \label{fig:bpqaoa128ap}
\end{subfigure}
\hspace{.09\linewidth}
\begin{subfigure}{0.3\linewidth}
\begin{subfigure}{\linewidth}
    \centering
    \includegraphics[width = .9\linewidth]{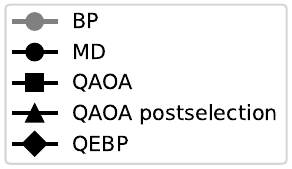}
    \phantomsubcaption
\end{subfigure}
\begin{subfigure}{\linewidth}
    \centering
    \raisebox{.4\linewidth}{\includegraphics[width = .7\linewidth]{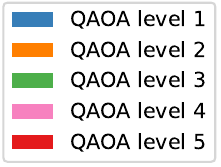}}
    \phantomsubcaption
\end{subfigure}
\phantomsubcaption
\end{subfigure}
\caption{Block error rate against signal-to-noise ratio $(E_0/N_b)$ for syndrome decoding with QAOA using a ranking system, QAOA with codeword post-selection and QAOA-enhanced belief propagation decoding. The test code is a $[12,8]$-regular code, and the simulations were run for $p = 1,\ldots, 5$.}
\label{fig:128ap}
\end{figure}

\Cref{fig:qaoa128ap} compares the BLER of the QAOA decoding algorithm, the sum-product algorithm, and the maximum-likelihood for the $(2,3)$-regular LDPC code. As shown in the figure, QAOA with five layers outperforms the best-known efficient classical decoding algorithm. However, by performing an additional post-selection step (see \cref{fig:qaoapost128ap}), the number of layers needed to outperform belief propagation is reduced to 2, achieving a lower block error rate than level 5 QAOA syndrome decoding. 

\Cref{fig:bpqaoa128ap} shows the BLER for the QAOA-enhanced belief propagation algorithm, comparing it to the sum-product algorithm and maximum-likelihood decoding, as before. QEBP with 1-layer QAOA gives an error performance similar to standard belief propagation.

\begin{figure}[ht]
\begin{subfigure}{.5\linewidth}
    \centering
      \includegraphics[width=\linewidth]{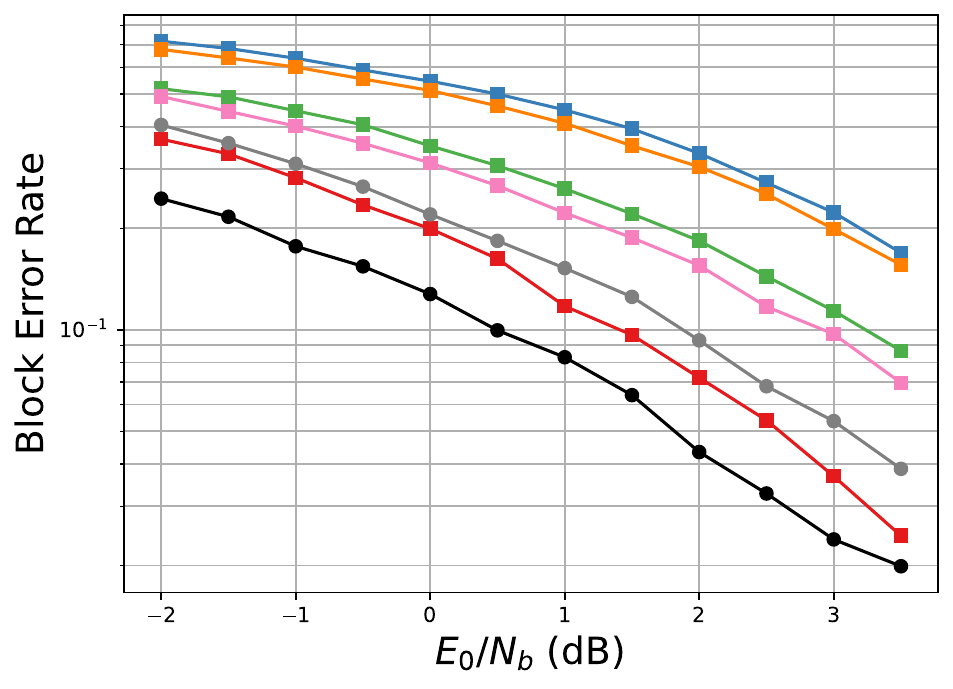}
      \caption{Syndrome decoding with QAOA.}
      \label{fig:qaoa128apirr}
\end{subfigure}%
\begin{subfigure}{.5\linewidth}
\centering
  \includegraphics[width=\linewidth]{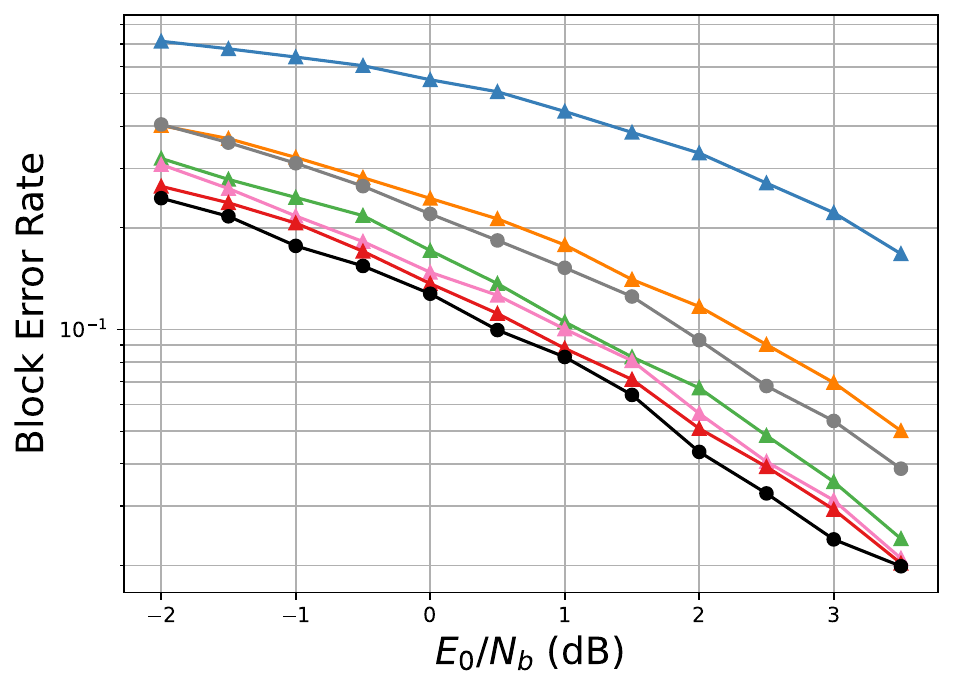}
  \caption{Decoding with codeword post-selection.}
  \label{fig:qaoapost128apirr}
\end{subfigure}\\[1ex]
\begin{subfigure}{.5\linewidth}
    \centering
    \includegraphics[width=\linewidth]{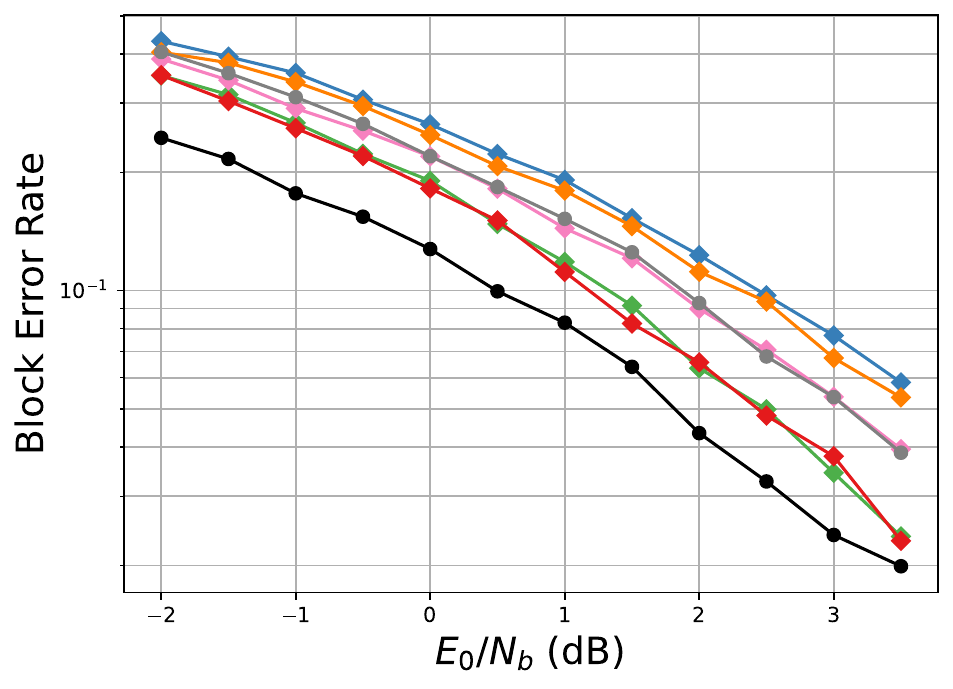}
  \caption{Decoding with QAOA-enhanced belief propagation.}
  \label{fig:bpqaoa128apirr}
\end{subfigure}
\hspace{.09\linewidth}
\begin{subfigure}{0.3\linewidth}
\begin{subfigure}{\linewidth}
    \centering
    \includegraphics[width = .9\linewidth]{plots/legend_shape.pdf}
    \phantomsubcaption
\end{subfigure}
\begin{subfigure}{\linewidth}
    \centering
    \raisebox{.4\linewidth}{\includegraphics[width = .7\linewidth]{plots/legend_colour.pdf}}
    \phantomsubcaption
\end{subfigure}
\phantomsubcaption
\label{fig:ber128apirr}
\end{subfigure}
\caption{Block error rate against signal-to-noise ratio $(E_0/N_b)$ for syndrome decoding with QAOA using a ranking system, QAOA with codeword post-selection and QAOA-enhanced belief propagation decoding. The test code is a $[12,8]$-irregular code, and the simulations were run for $p = 1, \ldots, 5$.}
\label{fig:128apirr}
\end{figure}

Similarly, the results for the $[12,8]$-irregular code are summarized in \cref{fig:128apirr}, which, as expected, has better error rate performance overall \cite{Luby2001}. In this case, all QAOA decoding algorithms are much closer in block error rate to minimum distance decoding than in the regular case. Remarkably, post-selecting for codewords in QAOA decoding gives a near maximum-likelihood performance for QAOA level 5.

The best-performing algorithm, irrespective of the number of QAOA layers, is the QAOA decoder with codeword post-selection. However, for level-1 QAOA, the QAOA-enhanced belief propagation approach either matches (\cref{fig:bpqaoa128ap}) or is very close (\cref{fig:bpqaoa128apirr}) to belief propagation.
    \section{Probability of error for the AWGN channel} \label{section:comm_channel}

Here, we summarize how we extract the probability of error from an AWGN channel with binary phase-shift key (BPSK) modulation, connecting it to the BSC channel. The encoded message is sent through a communication channel that may introduce bit-flip errors. We consider an additive white Gaussian noise (AWGN) channel with zero mean and variance $\sigma^2$. For the AWGN channel, the variance is given in terms of the noise power spectral density $N_0$ by
\begin{equation}
    \sigma^2 = \frac{N_0}{2},
\end{equation}
and the noise density function is
\begin{equation}
    f_N (n) = \frac{1}{\sqrt{2 \pi N_0/2}} \exp \left( \frac{-n^2}{N_0} \right).
\end{equation}

We use BPSK modulation, which segments bits into one tuple ($1 \rightarrow A$ and $0 \rightarrow -A$), which are then mapped into a waveform. Take $X$ to be the transmitted value of the waveform, $Y$ to be the received value, modified by the AWGN channel, and $\widehat{X}$ the assigned value after receiving $Y$. Using a Maximum A Posteriori (MAP) decoder, the resulting phase assignments are: $\widehat{X} = A$ if $Y> 0$ and $\widehat{X}=-A$ if $Y<0$. So, the probability of wrong detection is given by
\begin{equation}
    P(\widehat{X} \neq X) = \frac{1}{2}P(\widehat{X}=A|X=-A) + \frac{1}{2}P(\widehat{X}=-A|X=A) = P(N>A), 
\end{equation}
where $N \sim \mathcal{N}(0,N_0/2)$ is a Gaussian noise random variable corresponding to the AWGN channel. A measure of signal-to-noise ratio for a digital communication system is $E_b/N_0$, where $E_b$ is the energy per information bit. The probability of error is then related to $E_b/N_0$ by 
\begin{equation}
    P_e = \rm{Q}\left( \sqrt{\frac{2E_b}{N_0}}\right), \label{eq:perrorawgn}
\end{equation}
where $\rm{Q}$ is the q-function \cite{Gallager2006}. \Cref{eq:perrorawgn} gives the probability of error per bit. So, we can construct a BSC channel with crossover probability $\varepsilon = P_e$. 
\end{appendices}
\end{document}